\documentclass{emulateapj}
\usepackage{lscape}
\newcommand{\msun}{\mbox{M$_\odot$}}% Msun

\newcommand{\mean}[1]{\mbox{$\langle#1\rangle$}} %generic mean for defined qu.
\newcommand{\av}{\mbox{$A_V$}} % Visual Extinction

\shorttitle{1.1~mm Surveys of Serpens, Perseus, and Ophiuchus}
\shortauthors{Enoch et al.}

\begin{document}

\title{Comparing Star Formation on Large Scales in the c2d Legacy Clouds: 
Bolocam 1.1~mm Dust Continuum Surveys of Serpens, Perseus, and Ophiuchus}

\author{Melissa L. Enoch \altaffilmark{1}, Jason Glenn \altaffilmark{2}, Neal J. Evans II \altaffilmark{3}, 
Anneila I. Sargent \altaffilmark{1}, Kaisa E. Young \altaffilmark{3,4}, and Tracy L. Huard \altaffilmark{5}}

\email{MLE: menoch@astro.caltech.edu}
\altaffiltext{1}{Division of Physics, Mathematics \& Astronomy, California
Institute of Technology, Pasadena, CA 91125}
\altaffiltext{2}{Center for Astrophysics and Space
Astronomy, 389-UCB, University of Colorado, Boulder, CO 80309}
\altaffiltext{3}{The University of Texas at Austin, Astronomy Department, 
1 University Station C1400, Austin, TX, 78712-0259}
\altaffiltext{4}{Department of Physical Sciences, Nicholls State University,
Thibodaux, Louisiana 70301}
\altaffiltext{5}{Harvard-Smithsonian Center for Astrophysics, 60 Garden St., 
Cambridge, MA 02138}

\begin{abstract}
We have undertaken an unprecedentedly large 1.1~millimeter
continuum survey of three nearby star forming clouds using
Bolocam at the Caltech Submillimeter Observatory.   
We mapped the largest areas in each cloud at millimeter or submillimeter
wavelengths to date: 7.5 deg$^2$ in Perseus (Paper I), 10.8
deg$^2$ in Ophiuchus (Paper II), and 1.5 deg$^2$ in Serpens with a
resolution of 31\arcsec, detecting 122, 44, and 35 cores,
respectively.  Here we report on results of the Serpens survey and
compare the three clouds.   Average measured angular core sizes and
their dependence on resolution  suggest that many of the observed
sources are consistent with power-law  density profiles.   Tests of
the effects of cloud distance reveal that linear resolution strongly
affects measured  source sizes and densities, but not the shape of the
mass distribution.   Core mass distribution slopes in Perseus and
Ophiuchus ($\alpha=2.1\pm0.1$ and $\alpha=2.1\pm0.3$) are consistent
with recent measurements of the stellar IMF,  whereas the Serpens
distribution is flatter ($\alpha=1.6\pm0.2$).   We also compare the
relative mass distribution shapes to predictions from turbulent
fragmentation simulations.  Dense cores constitute less than 10\% of
the total cloud mass  in all three clouds, consistent with other
measurements of  low star-formation efficiencies.  Furthermore, most
cores are found at  high column densities; more than 75\% of 1.1~mm 
cores are associated with $\av \gtrsim 8$~mag in Perseus, 15~mag in 
Serpens, and $20-23$~mag in Ophiuchus.
\end{abstract}


\section{Introduction}

Large-scale physical conditions in molecular clouds influence the
outcome of local star formation, including the stellar initial mass
function, star-formation efficiency, and the spatial distribution of
stars within clouds \citep[e.g.][]{evans99}.   The physical processes
that  provide support of molecular clouds and control the
fragmentation of cloud material into star-forming cores remain
a matter of debate.  In the classical picture magnetic fields
provide support and collapse occurs via ambipolar diffusion
\citep[e.g.][]{shu78}, but many simulations now suggest that
turbulence dominates both support and fragmentation \citep[for a
  review see][]{mlk04}.  

Dense prestellar and protostellar condensations, or cores  \citep[for
  definitions and an overview see][]{dif05}, provide a  crucial link
between  the global processes that control star formation on large
scales and the properties of newly-formed stars.   The mass and spatial
distributions of such cores retain imprints  of the fragmentation
process, prior to significant influence from later  protostellar
stages such as mass ejection in outflows,  core dissipation, and
dynamical interactions.  These cold (10~K), dense ($n
>10^4$~cm$^{-3}$) cores are most easily observed at millimeter and
submillimeter wavelengths where continuum  emission from cold dust
becomes optically thin and traces the total  mass.   Thus, complete
maps of molecular clouds at millimeter wavelengths are  important for
addressing some of the outstanding questions in star formation.

Recent advances in millimeter and submillimeter wavelength continuum
detectors have enabled a number of large-scale surveys of nearby
molecular clouds \citep[e.g.][]{john04, kirk05, hatch05, enoch06,
  stanke06, young06}.  In addition to tracing the current and future
star-forming activity of the clouds on large scales, millimeter and
submillimeter observations are essential to understanding the
properties of starless cores and the envelopes of the most deeply
embedded protostars \citep[for more on the utility of  millimeter
  observations, see][]{enoch06}.

We have recently completed 1.1~mm surveys of Perseus
\citep[][hereafter Paper~I]{enoch06} and Ophiuchus  \citep[][hereafter
  Paper~II]{young06} with Bolocam at the Caltech Submillimeter
Observatory (CSO).  In this work we present a similar 1.1~mm survey of
Serpens, completing our three-cloud study of nearby northern
star-forming molecular clouds.  Unlike previous work, our surveys not
only cover the largest area in each cloud to date, but the uniform
instrumental properties allow a comprehensive comparison of the cloud
environments in these three regions.   A comparison of the results for
all three clouds provides insights into global cloud conditions and
highlights the influence  that cloud environment has on properties of
star forming cores.

Background facts on the Perseus and Ophiuchus molecular clouds are
discussed  in Papers I and II.  The Serpens molecular cloud is an
active star formation region at a distance of $d=260\pm10$~pc
\citep{stra96}.  Although the cloud extends more then 10~deg$^2$ as
mapped by optical extinction \citep{cam99}, most observations of the
region have been focused near the main Serpens cluster at a Right
Ascension (R.A.) of $18^{h}30^{m}$ and declination (decl.) of
$1\degr15\arcmin$ (J2000).   

The Serpens cluster is a highly extincted region with a high density
of young stellar objects (YSOs), including a number of Class~0
protostars.   It has been studied extensively at near-infrared,
far-infrared, submillimeter, and millimeter wavelengths
\citep[e.g.][]{ec92, hb96, lars00, davis99, ced93, ts98}.    Some
recent work has also drawn attention to a less well known cluster  to
the south, sometimes referred to as Serpens/G3-G6
\citep{djup06,harv06}.  Beyond these two clusters no continuum
millimeter or submillimeter continuum surveys have  been done that
could shed light on large-scale star formation processes.

Following Papers I and II, we utilize the wide-field mapping
capabilities of Bolocam, a large format bolometer array at the  CSO,
to complete millimeter continuum observations of 1.5~deg$^2$ of the
Serpens cloud.  These observations are coordinated to cover the area
mapped with  \textit{Spitzer} Space Telescope  IRAC and MIPS
observations of Serpens from the ``Cores to Disks''
\citep[c2d][]{eva03}  legacy project.  While millimeter and
submillimeter observations are essential to understanding the
properties of dense prestellar cores and protostellar envelopes,
infrared observations are necessary to characterize the protostars
embedded within those envelopes.   In a future paper (M. Enoch et al. 2007,
in preparation) we will take  advantage of the overlap between our
1.1~mm maps and the c2d  \textit{Spitzer} Legacy maps to characterize
the deeply embedded and prestellar populations in Serpens.

Here the Bolocam observations and data reduction (\S~\ref{obssect}),
general cloud morphology and source properties (\S~\ref{ressect}),
and summary (\S~\ref{serpsum}) for Serpens are briefly presented.  In
\S~\ref{compsect} we compare the millimeter survey results for
Serpens, Perseus, and Ophiuchus.   First, we outline our operational
definition of a millimeter core including  instrumental effects in
\S~\ref{coresect}, and discuss the observational  biases introduced by
different cloud distances in \S~\ref{convsect}.  Physical implications
of source sizes and shapes, and differences between the clouds, are
discussed in \S~\ref{sizecompsect}. We examine source densities  and
compare the cloud mass versus size distributions in
\S~\ref{dencompsect}.  We analyze the core mass distributions and
their relation to cloud turbulence  in \S~\ref{masscompsect}, and the
spatial distributions of the  core samples and clustering in
\S~\ref{clustsect}.  The relationship of dense cores to the surrounding 
cloud column density and the mass fraction in dense cores are discussed in
\S~\ref{threshsect} and \S~\ref{efficsect}.  
We end with a  summary in \S~\ref{sumsect}.

\section{Serpens Observations and Data Reduction}\label{obssect}

Observations and data reduction for Serpens follow the same
methodology  as for Perseus (Paper I) and Ophiuchus (Paper II).  The
data reduction techniques we have developed for the molecular cloud
data  are described in detail in Paper~I and Paper~II, including
removal of sky noise, construction of pointing and  calibration
models, application of iterative mapping, and source extraction.
Given these previous descriptions,  only details specific to Serpens
will be presented here.  Further information about the Bolocam
instrument and reduction pipeline is also available in
\citet{laurent05}.

\subsection{Observations}

As for Perseus and Ophiuchus, millimeter continuum observations of
Serpens  were made with
Bolocam\footnote{\url{http://www.cso.caltech.edu/bolocam}} at the CSO
on Mauna Kea, Hawaii.  Bolocam is a 144-element bolometer array that
operates at millimeter wavelengths and is designed to map large fields
\citep{glenn03}.   Observations of Serpens were carried out at a
wavelength of $\lambda=1.1$~mm, where the field of view is $7\farcm5$
and the beam is well approximated by a Gaussian with full-width at
half-maximum  (FWHM) size of 31$\arcsec$.  The instrument has a
bandwidth of 45 GHz at $\lambda=1.1$mm, excluding  the CO (J=2-1) line
to approximately 99\%.

\begin{figure}
\plotone{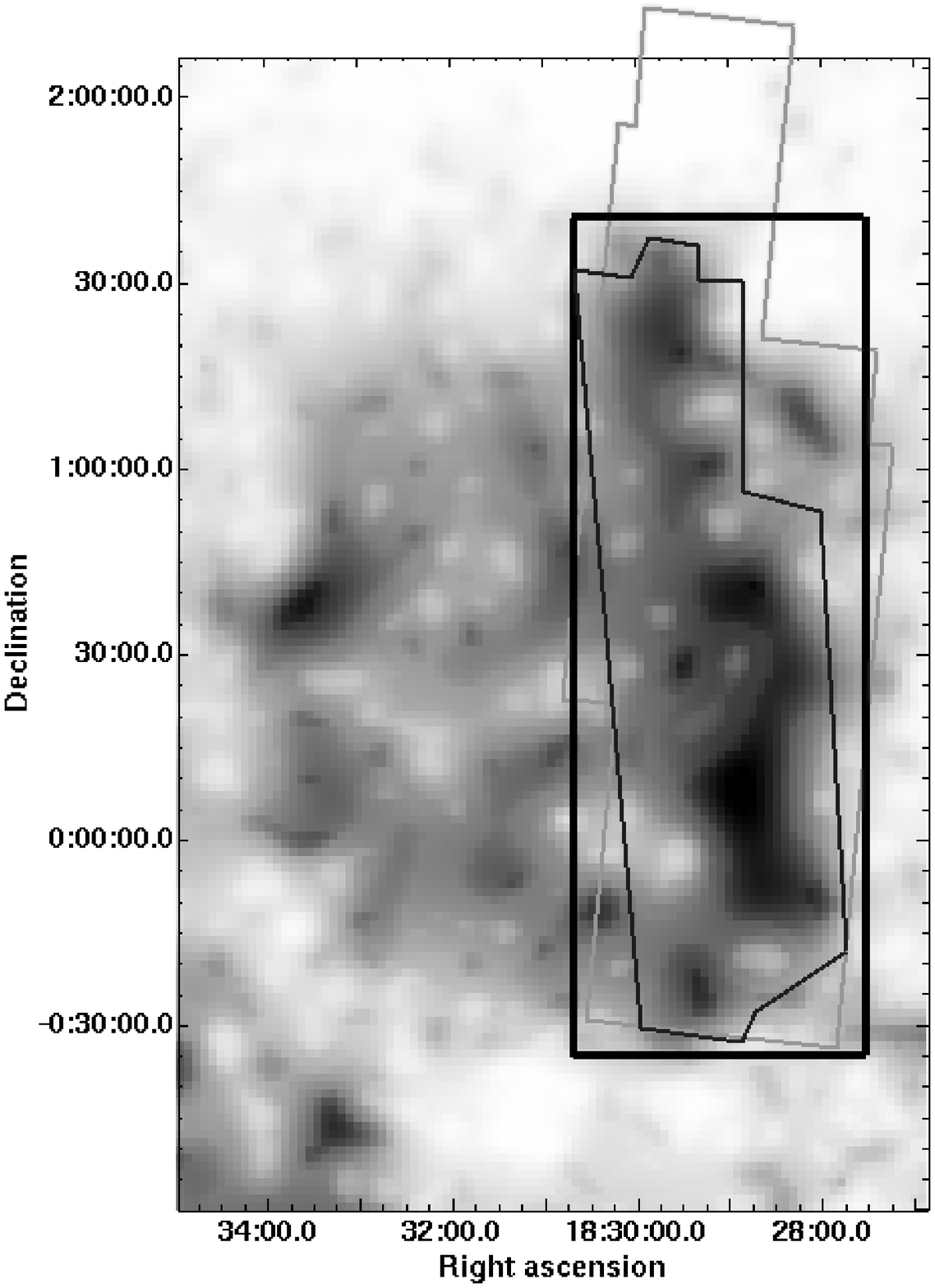}
\caption{Bolocam 1.1~mm (thick line) and Spitzer c2d IRAC (thin line)
and MIPS (gray line) coverage of Serpens overlaid on the \citet{cam99} 
visual extinction map.  The area observed with IRAC was chosen to cover 
$A_V \ge 6^m$ in this portion of the cloud.  Our Bolocam survey  covers 
the same area as the IRAC and a slightly smaller area than the MIPS 
observations.\label{c2dfig}}
\end{figure}

The 1.1mm observations were designed to cover a region with $\av \ge
6$~magnitudes ($\av \ge 6^m$) in the visual extinction  map of
\citet{cam99}, shown in Figure~\ref{c2dfig}.  As demonstrated in
Figure~\ref{c2dfig},  this ensures that the Bolocam observations
overlap as closely as possible with the region of Serpens observed
with \textit{Spitzer} IRAC and MIPS as part of the c2d Legacy project.
In  practice, the Bolocam survey covers a region slightly larger than
the IRAC map, and slightly smaller than the MIPS map.

Serpens was observed during two separate runs in 2003 May 21-June 9
and 2005 June 26-30.  During the 2003 run, 94 of the 144 channels
were operational, compared to 109 during the 2005 run.    Scans of
Serpens were made at a rate of $60\arcsec$~sec$^{-1}$ with no
chopping of the secondary.  The final map consists of 13 scans from
2003 in good weather, and 17 scans from 2005 in somewhat poorer
conditions.   Each scan covered the entire 1.5 deg$^2$ area and took
$35-40$ minutes to complete depending on the scan direction.   Scans
were made in two orthogonal directions, approximately half in R.A. 
and half in decl.  This strategy allows for good cross-linking in
the final map, sub-Nyquist sampling, and minimal striping from $1/f$
noise.  Scans were observed in sets of three offset by $-43\arcsec$,
$0\arcsec$, $+43\arcsec$ to optimize coverage.

In addition, small maps of pointing sources  were observed
approximately every 2 hours, and at least one primary flux
calibration source, including Neptune, Uranus, and Mars, was observed
each night.  Several larger beam maps of  planets were also made
during each run, to characterize the Bolocam beam at 1.1~mm.

\subsection{Pointing and Flux Calibration}\label{pointsect}

A pointing model for Serpens was generated using two nearby pointing
sources, G~34.3 and the quasar 1749+096.  After application of the
pointing model, a comparison to the literature SCUBA $850~\micron$
positions of four bright known sources \citep{davis99} in the main
Serpens cluster indicated a constant positional offset of
($\delta$R.A., $\delta$decl.) = (5$\arcsec$, $-10\arcsec$).  We
corrected for the positional offset, but estimate an uncertainty in
the absolute pointing of $10 \arcsec$, still small compared to the
beam size of 31\arcsec.   The relative pointing errors, which cause
blurring of sources and an increase in the effective beam size, should
be much smaller, approximately $5 \arcsec$.  Relative pointing errors
are characterized by the rms pointing uncertainty, derived from the
deviations of G~34.3 from the pointing model.
 
The Bolocam flux calibration method makes real-time corrections for
the atmospheric attenuation and bolometer operating point using the
bolometer optical loading, by calculating the calibration factor as a
function of the bolometer DC resistence.  Calibrator maps of Neptune,
Uranus, and G~34.3, observed at least once per night, were used to
construct a calibration curve for each run.  A systematic uncertainty
of approximately 10\% is associated with the absolute flux
calibration, but relative fluxes should be much more accurate.

\subsection{Cleaning and Iterative mapping}\label{cleansect}

Aggressive sky subtraction techniques are required for Bolocam data to
remove sky noise, which dominates over the astronomical signal before
cleaning.  As in Papers I and II, we remove sky noise from the Serpens
scans using Principal Component Analysis (PCA) cleaning
(\citet{laurent05} and references therein), by subtracting 3 PCA
components.

As described in Paper~I, PCA cleaning removes some source flux from
the map as well as sky noise, necessitating the use of an iterative
mapping procedure to recover lost astronomical flux.  In this
procedure, sky subtraction is refined by iteratively removing a source
model (derived from the cleaned map) from the raw data,  thereby
reducing contamination of the sky template by bright sources.
Simulations show that at least 98\% of source flux density is
recovered after 5 iterations, with the exception of very large
($\gtrsim 100\arcsec$ FWHM) faint sources, for which we only recover
90\% or less of the true flux density (See Paper~I).  We
conservatively estimate a 10\% residual photometric  uncertainty from
sky subtraction after iterative mapping, in addition to the 10\%
uncertainty in the absolute flux calibration.

Data from the 2003 and 2005 observing runs were iteratively mapped
separately because they required different pointing and flux
calibration models.  After iterative mapping the two epochs were
averaged, weighted by the square root of the observational  coverage.

\subsection{Source Identification}\label{idsect}

Millimeter cores are identified in an optimally filtered map using the
source extraction method  described in Paper~I.   Because the signal
from a point source lies in a limited frequency band, we can use an
optimal (Wiener) filter in Fourier space to attenuate $1/f$ noise at
low frequencies, as well as high frequency noise above the signal
band.  The optimal filter preserves the resolution of the map and the
peak  brightness of point sources but reduces the rms noise per pixel
by approximately $\sqrt{3}$, thereby optimizing the source
signal-to-noise (S/N).  Extended sources will have slightly enhanced
peak values in the optimally filtered map.  After optimal filtering,
the map is trimmed to remove areas of low coverage.  Note that the
optimally filtered map is used for source detection only; all
photometry is measured in the unfiltered map, and all maps displayed
here are unfiltered.

Observational coverage, which depends on the scan strategy, number  of
scans, and number of bolometer channels, was very uniform for Serpens;
trimming regions where the coverage was less than 30\% of the peak
coverage was equivalent to cutting off the noisy outer edges of the
map.  The average coverage of the Serpens map is 1600 hits per pixel,
where a hit means a bolometer passed over this position, with
differences in hits per pixel across the map of 18\%.  The average
coverage corresponds to an integration time of 13 minutes per pixel,
although individual pixels  are not independent because the map is
over-sampled. 

For each pixel, the local rms noise is calculated in small (45~
arcmin$^2$) boxes using a noise map from which sources have been
removed.  This noise map is derived as part of the iterative mapping
process (see Paper~I).   A simple peak-finding routine identifies all
pixels in the optimally filtered map more than $5\sigma$ above the
local rms noise level, and source positions are determined using an
IDL centroiding routine.  Each new source must be separated by at
least a beam size from any previously identified source centroid.  The
centroid is a weighted average position based on the surface
brightness within a specified aperture, and is computed as the
position at which the derivatives of the partial sums of the input
image over (y,x) with respect to (x,y) equal zero.  A given centroid
is considered ``well defined'' as long as the computed derivatives are
decreasing.  All sources were additionally inspected by eye to remove
spurious peaks near the noisier edges of the map.

\section{Serpens Results}\label{ressect}

\begin{figure}
\notetoeditor{Figure 2 should be full-page.}
\plotone{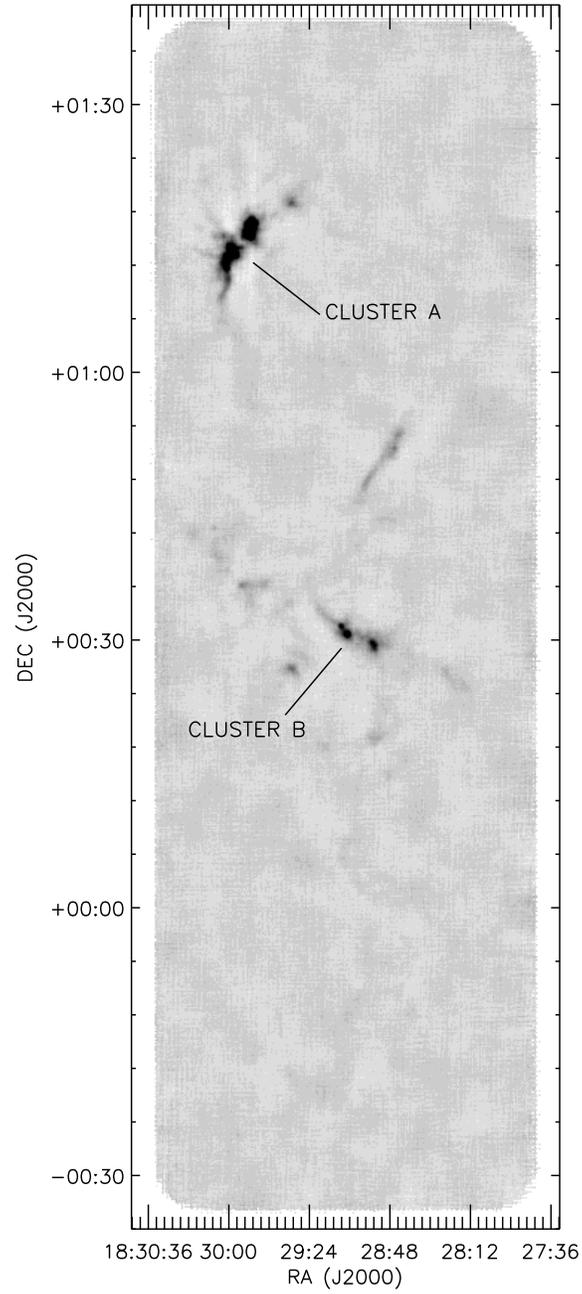}
\caption{Bolocam 1.1~mm map of 1.5~deg$^2$ (31~pc$^2$ at $d=260$~pc)
in the Serpens molecular cloud.  Bolocam has a resolution of 
$31\arcsec$, and the map is binned to 10$\arcsec$~pixel$^{-1}$.
The average $1\sigma$ rms noise is 9.5~mJy beam$^{-1}$.
\label{mapfig}}
\end{figure}

The final $10\arcsec$ pixel$^{-1}$ Serpens map is shown in
Figure~\ref{mapfig}, with the well known northern Serpens cluster,
Cluster~A \citep{harv06}, indicated, as well as the  southern cluster,
Cluster B (Serpens/G3-G6).   Covering a total area of 1.5  deg$^2$, or
30.9~pc$^2$ at a distance of 260~pc, the map has a linear resolution
of $7.3\times10^3$~AU. 

\begin{figure}
\plotone{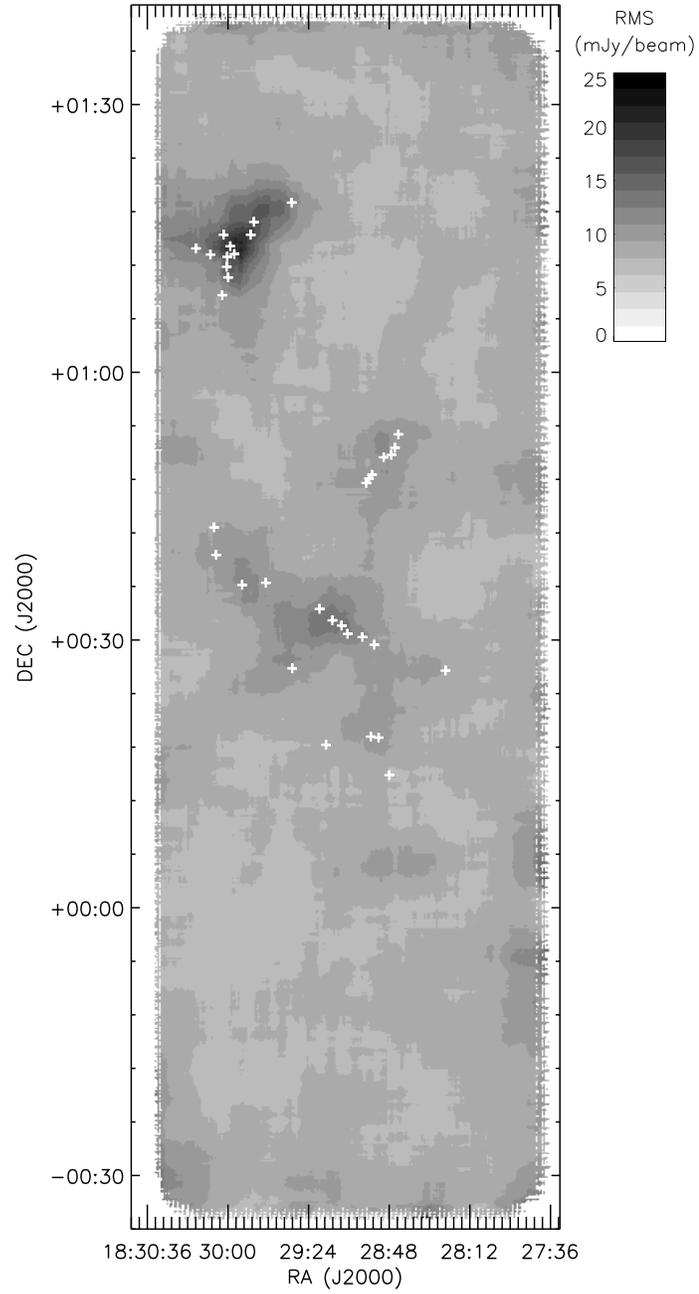}
\caption{Map of the local 1$\sigma$ rms noise per beam in the Serpens
  map.  Positions of the 35 sources are indicated by white plus
  symbols.   The average rms noise is 9.5  mJy~beam$^{-1}$, varying by
  18\% across  the map. The noise is higher  near bright sources due
  to sky subtraction residuals.  \label{noisefig}}
\end{figure}

We identify 35 sources above the $5\sigma$ detection limit in the
Serpens map.  The $5\sigma$ limit is based on the local rms noise, and
is typically  50 mJy beam$^{-1}$.  The noise map for Serpens is shown
in Figure~\ref{noisefig},  with the positions of identified sources
overlaid.  The mean rms noise is 9.5 mJy~beam$^{-1}$, but is higher
near bright sources.  Most of the 18\% variations in the local rms
noise occur in the main cluster region, where calculation of the noise
is confused by residual artifacts from bright sources.  Such artifacts
must contribute significantly to noise fluctuations; coverage
variations alone would predict rms noise variations of only $\sqrt{18}
= 4\%$.  This means that faint sources near bright regions have a
slightly lower  chance of being detected than those in isolation.

\begin{figure}
\plotone{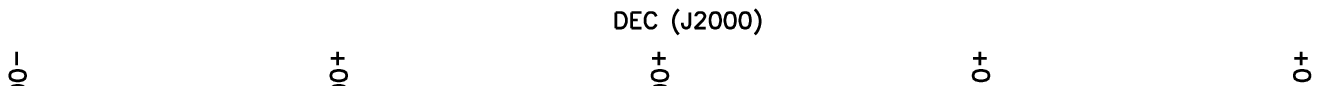}
\caption{Bolocam 1.1~mm map of Serpens with the positions of the 35
  sources detected above $5\sigma$ indicated by red circles.  Inset
  maps magnify the most densely populated source regions, including
  the well known northern Cluster A, Cluster B to the south, and an
  elongated filament reminiscent of the B1 ridge in Perseus (Paper~I).
  Despite the low rms noise level reached (9.5 mJy~beam$^{-1}$), few
  sources are seen outside the cluster regions.  Many bright 1.1~mm
  sources are associated with  YSOs \citep{harv06}, and all are
  coincident with 160~$\micron$ emission.
\label{sourcefig}}
\end{figure}

Source positions are listed in Table~\ref{srctab}, and identified by
red circles in Figure~\ref{sourcefig}.  Figure~\ref{sourcefig} also
shows magnifications of the more densely populated source regions,
including Cluster A and Cluster B.  We do not see any circularly
symmetric extended emission on scales $\gtrsim 3\arcmin$ in the map.
It should, in principle, be possible to recover symmetric structures
up to the array size of $7\farcm5$, but our simulations show that
sources $\gtrsim 4\arcmin$ in size are severely affected by cleaning
and therefore difficult to fully recover with iterative mapping.  The
map does contain larger filamentary structures up to $8\arcmin$ long.
In particular, the long filament between Cluster A and Cluster B is
reminiscent of the elongated ridge near B1 in Perseus (Paper~I).  The
Serpens filament does not contain the bright compact sources at either
end that are seen in the B1 ridge, however.

Previous millimeter-wavelength maps of Cluster A, such as the 1.1~mm
UKT14 map of \citet{ced93} and the $850~\micron$ SCUBA map of
\citet{davis99}, generally agree with our results in terms of
morphology and source structure.  Not all of the individual
$850~\micron$ sources are detected by our peak finding routine,
presumably due to the poorer resolution of Bolocam ($30\arcsec$)
compared to SCUBA ($14\arcsec$), but most can, in fact, be identified
by eye in the Bolocam map.  An IRAM 1.3~mm continuum map of  Cluster B
with 11\arcsec\ resolution \citep{djup06} is visually quite similar to
our Bolocam map of the region.  We detect each of the four 1.3~mm
sources identified (MMS1-4), although the 1.3~mm triplet MMS1 is seen
as a single extended source in our map.

Most of the brightest cores, in particular those in Cluster A, are
associated with known YSOs including a number of Class~0 objects
\citep{hb96,harv06}.   All bright 1.1~mm sources are aligned with
bright $160~\micron$ emission in the \textit{Spitzer} MIPS map of
Serpens observed by the c2d legacy project \citep{harv07}.  Fainter
millimeter sources are usually associated with extended $160~\micron$
filaments, but do not necessarily correspond to point sources in the
MIPS map.  Conversely, one bright extended region of $160~\micron$
emission just south of  Cluster A contains no 1.1~mm sources.   This
area also exhibits extended emission at 70 and 24~$\micron$, which may
be indicative of warmer, more diffuse material than  that of the dense
cores detected at 1.1~mm.  

\begin{figure}
\notetoeditor{Figure 5 should be full-page if possible.}
\plotone{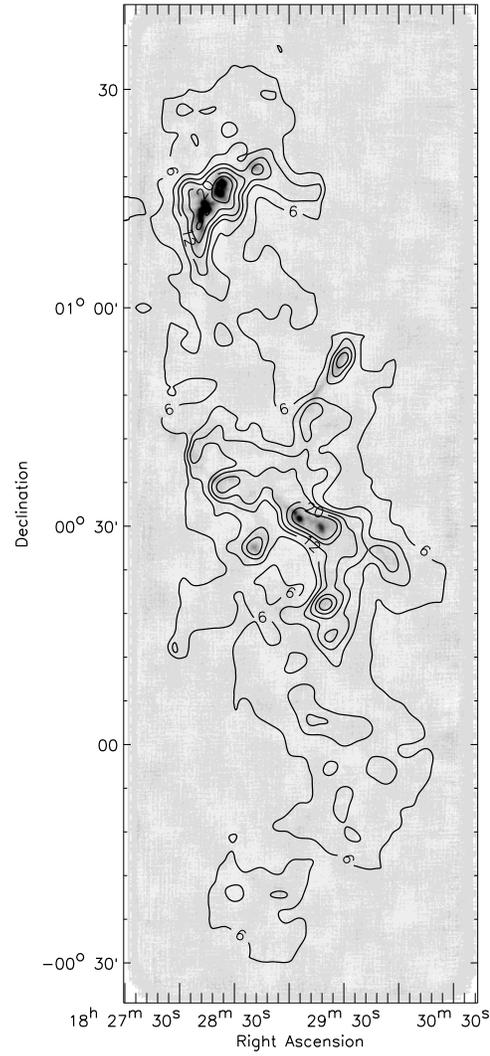}
\caption{Visual extinction (\av) contours calculated from c2d Spitzer
  maps, as described in \S~\ref{masssec}, overlaid on the gray-scale
  1.1~mm map.  Contours are $\av= 6,9,12,15,20,25^m$ and are smoothed
  to an effective resolution of $2\farcm5$.  All bright 1.1~mm cores
  are found in regions of high ($>15^m$) extinction, but not all high
  \av\ areas are  associated with strong millimeter sources. 
\label{avfig}}
\end{figure}

Despite the low rms noise level achieved in Serpens, very few sources
are seen outside the main clusters; most of the area that we mapped
appears devoid of 1.1~mm emission, despite being in a region of high
extinction.   Figure~\ref{avfig} shows a comparison between the
Bolocam millimeter map (gray-scale) and visual extinction (contours)
derived from c2d near- and mid-infrared Spitzer data.  

The majority of sources (90--95\%) detected by IRAC and MIPS in the
c2d clouds  have spectral energy distributions characteristic of
reddened stars.  Thus we  have measures of the visual extinction for
many lines of sight through the molecular clouds imaged by c2d.
Line-of-sight extinction values are derived by fitting the R$_V$=5.5
dust model   of \citet{wd01} to the near-infrared through mid-infrared
SED  \citep{deliv}.  For each of the three clouds,  the derived
line-of-sight extinctions were convolved with uniformly  spaced
90\arcsec\ Gaussian beams to construct an  extinction map.

Extinction maps used throughout this paper for Serpens, Perseus, and
Ophiuchus are derived from c2d data by this method.  The c2d
extinction  maps accurately trace column densities up to $\av \sim
40^m$,  but are relatively insensitive to small regions of high
volume density,  because they rely on the detection of background
stars.  Thus the extinction  maps are complementary to the 1.1~mm
observations, which trace high volume  density structures (see
\S~\ref{coresect}).  In Figure~\ref{avfig}, the \av\ map for Serpens
is smoothed to an  effective resolution of $2\farcm5$.

As can be seen in Figure~\ref{avfig}, nearly all Serpens millimeter
sources lie within regions of  high visual extinction ($\av \ge 10^m$)
and, in particular, all bright Bolocam sources are associated with
areas of $\av \ge 15^m$. Nevertheless, there are a number of high
extinction areas ($\av  \ge 12^m$) with no detectable 1.1~mm sources.
A similar general trend was noted in both Perseus (Paper~I) and
Ophiuchus (Paper~II), with relatively few sources found outside the
major groups and clusters associated with the highest extinction.
\S~\ref{threshsect} examines the relationship between $A_V$ and 1.1~mm
sources in more detail.

\subsection{Source Properties}\label{statsect}

\subsubsection{Positions and Photometry}

Positions, peak flux densities, and S/N for the 35 1.1~mm sources
identified in the Bolocam map of Serpens are listed in
Table~\ref{srctab}.    The S/N ratio is measured in the optimally
filtered map, whereas photometry and all other source properties are
measured in the unfiltered, surface brightness normalized map.   The
peak flux density per beam ($I_{\nu}$) is given in mJy beam$^{-1}$  
(1~mJy beam$^{-1} = 0.04$~MJy~sr$^{-1}$).  Uncertainties in
photometry are calculated from the local rms beam$^{-1}$,   calculated
as in \S~\ref{idsect}.   An additional systematic error of 15\% is
associated with all flux densities, from the absolute calibration
uncertainty and the systematic bias remaining after iterative mapping.
Table~\ref{srctab} also lists the most commonly used name from the
literature for known sources, and indicates if the 1.1~mm source is
coincident (within $60\arcsec$) with a MIPS $24~\micron$ source from
the c2d database \citep{harv07}.

Table~\ref{phottab} lists photometry in fixed apertures  of diameter
$40\arcsec$, $80\arcsec$, and $120\arcsec$, and the total integrated
flux density ($S_{\nu}$).   Integrated flux densities are measured
assuming a sky value of zero, and include a correction for the
Gaussian beam so that a point source has the same integrated flux
density in all apertures.   No integrated flux density is given if 
the distance to the nearest neighboring source is smaller than the
aperture diameter.   The total flux density is integrated in the
largest aperture ($30\arcsec-120\arcsec$ diameters in steps of
$10\arcsec$) that is smaller than the distance to the nearest
neighboring source.  Uncertainties are $\sigma_{ap} = \sigma_{mb}
(\theta_{ap}/\theta_{mb})$, where $\sigma_{mb}$ is the local rms
beam$^{-1}$ and ($\theta_{ap}$, $\theta_{mb}$) are the aperture and
beam FWHM respectively.

\begin{figure}
\plotone{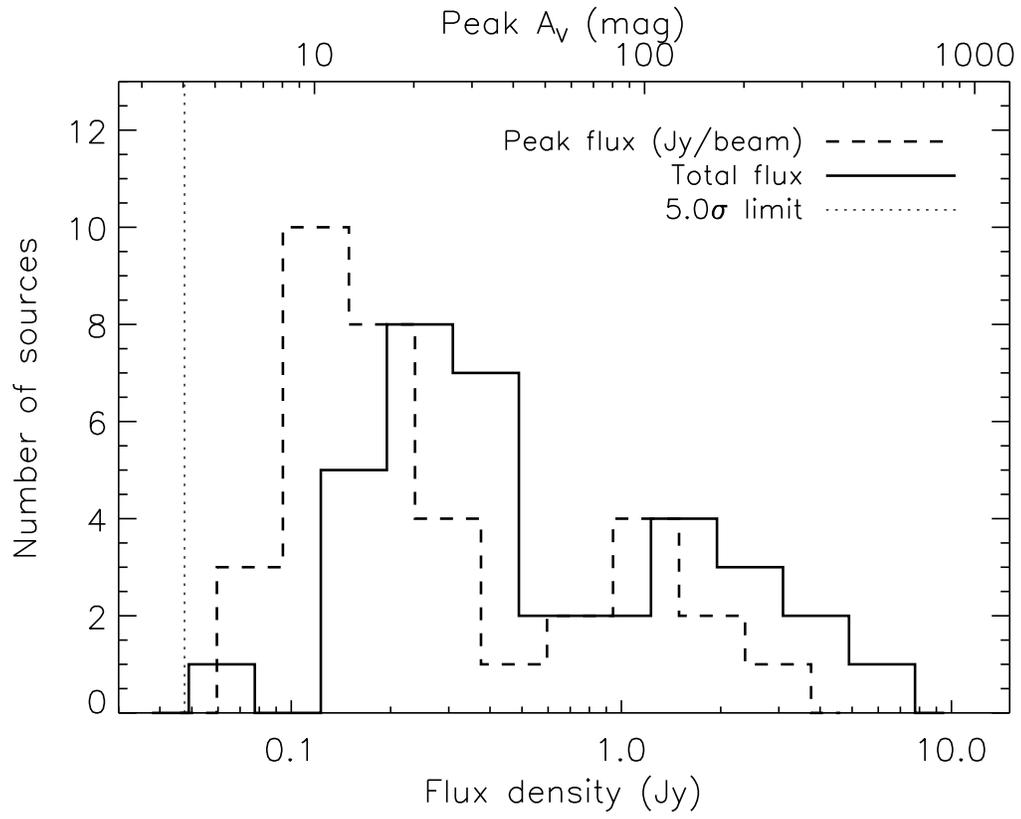}
\caption{Distribution of peak (dashed line) and total (solid line)
  flux densities of the 35 1.1~mm sources in Serpens.  Peak \av\ values 
  derived from the 1.1~mm peak flux densities using Eq.~\ref{aveq} are 
  shown on the upper axis.  The mean peak flux density of the sample
  is $0.5$~Jy~beam$^{-1}$, the mean peak \av\ is 40$^m$, and the
  mean total flux density is 1.0~Jy.  The $5\sigma$ detection limit
  of 0.05~Jy (dotted line) is relatively uniform across the
  cloud. \label{fluxfig}}
\end{figure}

Peak and total flux density distributions for the 35 1.1~mm sources 
in Serpens are shown in Figure~\ref{fluxfig}, with the $5\sigma$
detection limit indicated.  In general, source total flux densities
are larger than peak flux densities because most sources in the map
are extended, with sizes larger than the beam.  Both distributions
look bimodal; all of the sources in the brighter peak are in either
Cluster A or Cluster B.  The mean peak flux density of the sample is
0.5~Jy beam$^{-1}$, and the mean total flux density is 1.0~Jy, both
with large standard deviations of order the mean value.  Peak
\av\ values of the cores, calculated from the peak flux density as in
\S~\ref{masssec}, are indicated on the upper axis.

\subsubsection{Sizes and Shapes}

Source FWHM sizes and position angles (PA, measured east of north)
are measured by fitting an elliptical Gaussian after masking out
nearby sources using a mask radius equal to half the distance to the
nearest neighbor.  The best fit major and minor axis sizes and PAs are
listed in Table~\ref{phottab}.   Errors given are the formal fitting
errors and do not include uncertainties due to residual cleaning
effects, which are of order $10-15\%$ for the FWHM and $5\degr$ for
the PA (see Paper~I).

\begin{figure}
\plotone{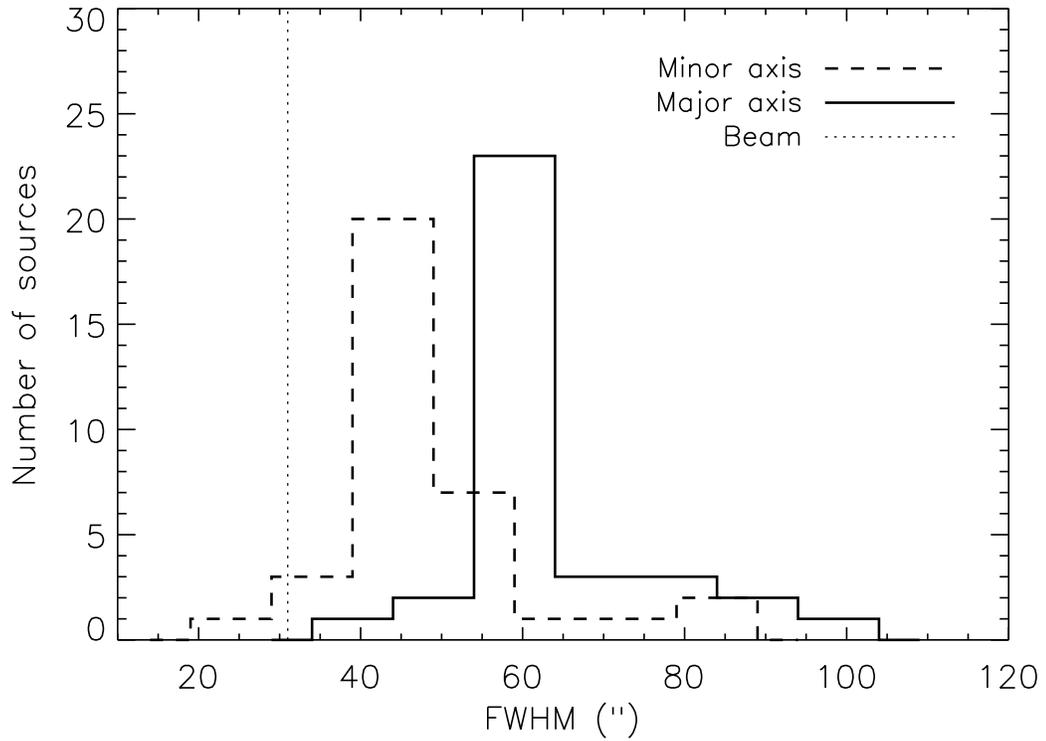}
\caption{Distribution of source FWHM minor axis (dashed line) and
  major axis (solid line) sizes, determined by an elliptical gaussian
  fit.  The mean FWHM sizes are $49\arcsec$ (minor axis) and
  $63\arcsec$ (major axis), and the mean axis ratio (major/minor) is
  1.3. \label{sizefig}}
\end{figure}

As can be seen in Figure~\ref{sizefig}, the minor axis FWHM values are
fairly narrowly distributed around the sample mean of $49\arcsec$,
with a standard deviation of $12\arcsec$.  The major axis FWHM have a
similarly narrow distribution with a mean of $63\arcsec$ and a scatter
of $12\arcsec$.  On average sources are slightly elongated, with a
mean axis ratio at the half-max contour (major axis FWHM / minor axis
FWHM) of 1.3.

A morphology keyword for each source is also given in
Table~\ref{phottab}, to describe the general source shape and
environment.  Keywords indicate if the source is multiple (within
$3\arcmin$ of another source), extended (major axis full-width at
$2\sigma > 1\arcmin$), elongated (axis ratio at $4\sigma > 1.2$),
round (axis ratio at $4\sigma < 1.2$), or weak (peak flux densities
less than 5 times the rms per pixel in the unfiltered map).  The
majority (28/35) of sources are multiple by this definition, and
nearly all (32/35) are extended at the $2\sigma$ contour.

\subsubsection{Masses, Densities, and Extinctions}\label{masssec}

The total mass $M$ of gas and dust in a core is proportional to the
total flux density $S_{\nu}$, assuming the dust emission at 1.1~mm is
optically thin and both the dust temperature and opacity are
independent of position within a core:
\begin{equation}
M = \frac{d^2 S_{\nu}}{B_{\nu}(T_D) \kappa_{\nu}}, \label{masseq}
\end{equation}
where $\kappa_{1.1mm}=0.0114$~cm$^2$~g$^{-1}$ is the dust opacity per
gram of gas, $d=260$~pc is the distance, and $T_D=10$~K is the dust
temperature.  A gas to dust mass ratio of 100 is included in
$\kappa_{1.1mm}$.  The dust opacity is interpolated from
\citet[][Table~1 column 5]{oh94},  for dust grains with thin ice
mantles, coagulated for $10^5$ years at a gas density of
$10^6$~cm$^{-3}$.  This opacity has been found to be the best fit in a
number of radiative transfer models \citep{evans01, shir02, young03}.  

Masses calculated in this manner, assuming a dust temperature of
$T_D=10$~K for all sources, are listed in Table~\ref{phottab}.
Uncertainties given are from the uncertainty in the total flux density
only; additional uncertainties from $\kappa$, $T_D$, and $d$ together
introduce a total uncertainty in the mass of up to a factor  of 4 or
more (also see Paper~I).    The dust opacity is uncertain by up to a
factor of two or more  \citep{oh94}, owing to large uncertainties in
the assumed dust  properties, and possible positional variations of
$\kappa_{\nu}$  within a core.  Variations in both $\kappa_{\nu}$ and
the cloud distance $d$ have  smaller effects than dust temperature
uncertainties for the range  of plausible values
($\kappa_{1.1mm}=0.005-0.02$~cm$^2$~g$^{-1}$;  $d=200-300$~pc;
$T_D=5-30~K$).

Assuming $T_D =10$~K is a reasonable compromise to cover both
prestellar and protostellar sources, based on the results of more
detailed radiative transfer models \citep[][see also discussion in
  Paper I]{evans01,shir02}.  A temperature of 10~K will result in
overestimates of the masses of protostellar cores, which will be
warmer on the inside, by up to a factor of three (for a 20~K source).
Temperature errors are not as problematic as they might seem, however,
as most of the envelope mass is located at large radii and low
temperatures (see also Paper I).  

The total mass of the 35 1.1~mm cores in Serpens is $92\msun$, which
is only 2.7\% of the total cloud mass  (3470 \msun).  The cloud mass
is estimated  from the c2d visual extinction map using
$N($H$_2)/\av\ = 0.94 \times 10^{21}$~mag~cm$^{-2}$ \citep{bohlin78}
and
\begin{equation}
M\mathrm{(cloud)} = d^2 m_H \mu_{H_2} \Omega
\sum{N(\mathrm{H}_2)}\label{avmasseq}
\end{equation}
where $d$ is the distance, $\Omega$ is the solid angle, $\sum$
indicates  summation over all $\av>2^m$ pixels in the extinction map,
and $ \mu_{H_2}=2.8$ is the mean molecular weight per H$_2$ molecule.

\begin{figure}
\plotone{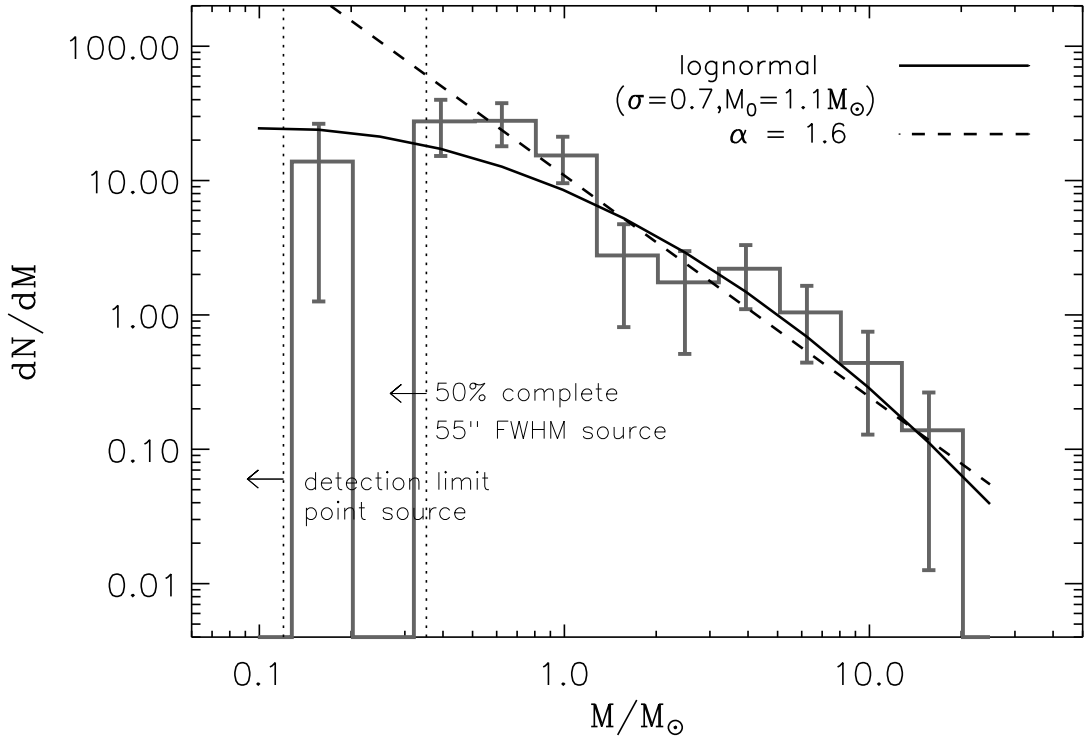}
\caption{ Differential mass distribution of the 35 detected 1.1~mm
  sources in Serpens for  masses calculated with $T_D = 10$~K.
  Dotted lines indicate the point source detection limit and the
  empirically derived 50\% completeness limit for sources with the
  average  FWHM size of $55\arcsec$.  The best fit power law ($\alpha
  = 1.6\pm0.2$) is shown, as well as the  best fit lognormal
  function. \label{massfn}}
\end{figure}

Figure~\ref{massfn} shows the differential mass function of all 1.1~mm
sources in Serpens.  The point source detection limit of $0.13
M_{\sun}$ is indicated, as well as the 50\% completeness limit for
sources of FWHM 55$\arcsec$ ($0.35 M_{\sun}$), which is the average
size of the sample.  Completeness is determined from Monte Carlo
simulations of simulated sources inserted into the raw data and run
through the reduction pipeline, as described in Paper~I.  The best fit
power law slope to $dN/dM \propto M^{-\alpha}$ is shown ($\alpha =
1.6$),  as well as the best fit lognormal slope.

Our mass distribution has a flatter slope than that found by
\citet{ts98} from higher resolution ($5\arcsec$) OVRO observations
($\alpha = -2.1$ for $M>0.35\msun$).   This may be due in part to the
fact that most of our  detections, at least 25/35, lie outside the
$5\farcm 5 \times 5\farcm 5$ area observed by \citet{ts98}.
The resolution differences of the observations may also contribute
significantly; for example, a number of our bright sources break down
into  multiple objects in the $5 \arcsec$ resolution map.  

From the peak  flux density $I_{\nu}$ we calculate the central H$_2$
column density for each source:
\begin{equation}
N(\mathrm{H}_2) = \frac{I_{\nu}}{\Omega_{mb} \mu_{H_2} m_H
  \kappa_{\nu} B_{\nu}(T_D)}. \label{aveq}
\end{equation}
Here $\Omega_{mb}$ is the beam solid angle, $m_H$ is the mass of
hydrogen, $\kappa_{1.1mm} = 0.0114$~cm$^2$g$^{-1}$ is the dust opacity
per gram of gas, $B_{\nu}$ is the Planck function, $T_D$=10~K is the
dust temperature, and $\mu_{H_2}=2.8$ as above.  From column density
we convert to  extinction:  $\av = N($H$_2)/ 0.94 \times
10^{21}$~mag~cm$^{-2}$ \citep{bohlin78}, adopting $R_V = 3.1$.  We
note, however, that this relation was determined for the diffuse
interstellar medium and may not be ideal for the highly extincted
lines of sight probed here.  

The resulting peak \av\ values are listed in Table~\ref{phottab}.
The average peak \av\ of the sample is 41$^m$ with a large standard
deviation of 55$^m$ and a maximum \av\ of 256$^m$.  Extinctions
calculated from the millimeter emission are generally higher than
those from the c2d visual extinction map by  approximately a factor of
7, likely a combination of both the higher  resolution of the Bolocam
map  ($30''$ compared with $90''$) and the fact that the extinction
map cannot trace the highest volume densities  because it relies on
the detection of background sources.  Grain growth in dense cores
beyond that included in the dust opacity from \citet{oh94}  could also
lead to an overestimate of the \av\ from out 1.1~mm data.

Also listed in Table~\ref{phottab} is the mean particle density:
\begin{equation}
\mean{n} = \frac{M}{\frac{4}{3} \pi R^3 m_H \mu_p}, \label{deneq}
\end{equation}
where $M$ is the total mass, $R$ is the linear deconvolved half-width
at half-max size, and $\mu_p=2.33$ is the mean molecular weight per
particle.  The median of the source mean densities is
$2.3\times10^5$~cm$^{-3}$, with  values ranging from $3.1\times10^4$
to $6.1\times10^6$~cm$^{-3}$.

\section{Serpens Summary}\label{serpsum}

We have completed a 1.1~mm dust continuum survey of Serpens, covering
$1.5$~deg$^2$, with Bolocam at the CSO.   We identify 35 1.1~mm
sources in Serpens above a $5\sigma$ detection limit which is
50~mJy~beam$^{-1}$, or $0.13\msun$, on average.   The sample has an
average mass of $2.6\msun$, and an average source FWHM size of
$55\arcsec$.  On average, sources are slightly elongated with a mean
axis ratio at half-max of 1.3.  The differential mass distribution of
all 35 cores is consistent with a power law of slope $\alpha = 1.6\pm
0.2$ above $0.35 \msun$.  The total mass in dense 1.1~mm cores in
Serpens is $92\msun$, accounting for  2.7\% of the total cloud mass,
as estimated from our c2d visual extinction map.

\section{Three-Cloud Comparison of Perseus, Ophiuchus, and Serpens}\label{compsect}

The survey of Serpens completes a three-cloud study  investigating the
properties of millimeter emission in nearby star-forming molecular
clouds:  Perseus \citep{enoch06}, Ophiuchus \citep{young06}, and
Serpens.   Having presented the results for the Serpens cloud, we now
compare the three clouds.

Our large-scale millimeter surveys of Perseus, Ophiuchus, and Serpens,
completed with the same instrument and reduction techniques, provide
us with a unique basis for comparing the properties of 1.1~mm emission
in a variety of star-forming  environments.  In the following sections
we examine similarities and  differences in the samples of
star-forming cores, and discuss implications for physical properties
of cores as well as global cloud conditions.

\subsection{What is a core?}\label{coresect}

Before comparing results from the three clouds, we first describe our
operational definition of a millimeter core.  The response of Bolocam
to extended emission, together with observed sensitivity limits
determine the type of structure that is detectable in our 1.1~mm maps
The Bolocam 1.1~mm observations presented here are sensitive to
sub-structures in molecular clouds with volume density $n\gtrsim 2
\times 10^4$~cm$^{-3}$.   One way to see this is to calculate the mean
density along the curve  defined by the detection as a function of
size in each cloud.

\begin{figure}
\plotone{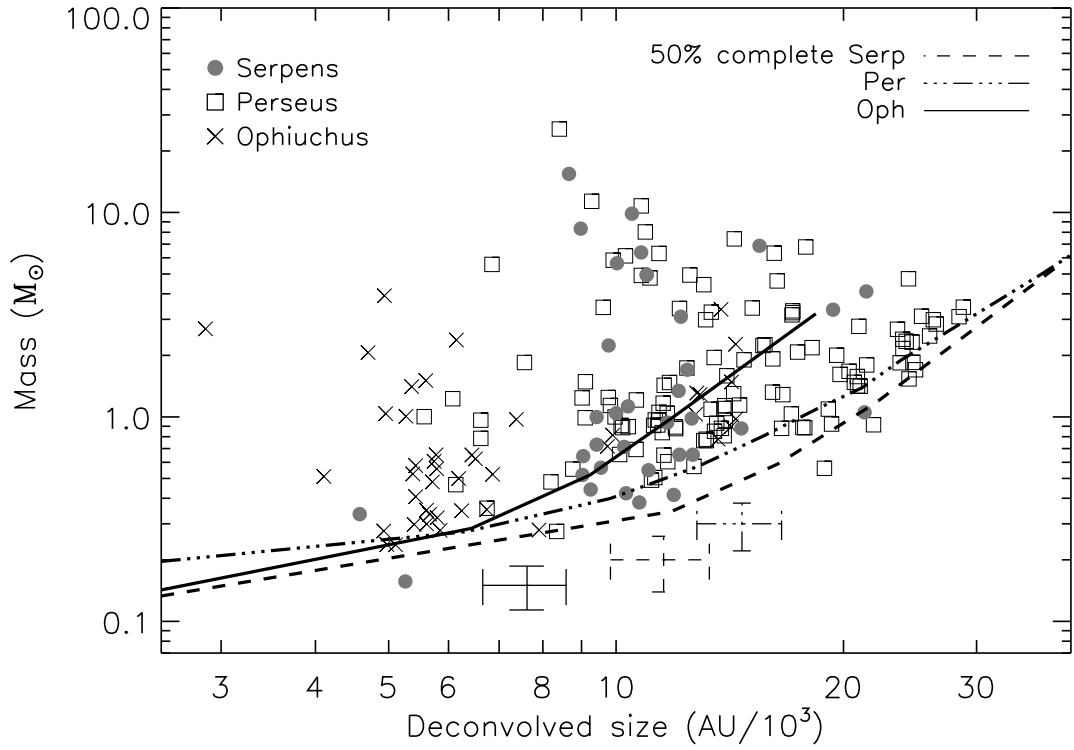}
\caption{Completeness as a function of linear deconvolved source size
  in  Serpens, Perseus, and Ophiuchus.  Symbols show the distribution
  of source mass versus deconvolved size in each of the three clouds,
  where the size is the linear deconvolved average FWHM.  Curves are
  empirical 50\% completeness limits determined from Monte Carlo
  simulations, and demonstrate the dependence of completeness  on
  source size and cloud distance.  
  The beam FWHM of $31''$ corresponds to approximately $8\times 10^3$~AU 
  in Serpens and Perseus, and $4\times10^3$~AU in Ophiuchus.
  Error bars for average sized
  sources near the detection limit in each cloud are also shown, as
  estimated from the results of Monte Carlo simulations and  pointing
  uncertainties of approximately $10\arcsec$.  
\label{mvscomp}}
\end{figure}

Figure~\ref{mvscomp} demonstrates how the completeness in each cloud
varies as a function of source size.  Plotted symbols give  the total
mass versus linear deconvolved size for all sources  detected in each
cloud, and  lines indicate the empirically derived 50\% completeness
limits.    For Ophiuchus the average completeness curve is plotted; as
the rms noise varies considerably in Ophiuchus, some regions have
higher or lower completeness limits than the curve shown here.
Completeness is determined from Monte Carlo simulations by adding
simulated sources to the raw data, processing them in the same way as
the real data, and attempting to detect them using our peak-finding
algorithm (see Paper~I).  We are biased against detecting large
diffuse  sources because we detect sources based on their peak flux
density, whereas the mass is calculated from the total flux, which
scales approximately as the size squared.

Calculating mean densities along the 50\% completeness curve in each
cloud yields $n_{lim}\sim 3-7\times 10^4$~cm$^{-3}$ in Serpens,
$n_{lim}\sim 2-4\times 10^4$~cm$^{-3}$ in Perseus, and $n_{lim}\sim
10-30\times 10^4$~cm$^{-3}$ in Ophiuchus.  By comparison, the mean
cloud density as probed by the extinction map is approximately
1000~cm$^{-3}$ in Serpens, 220~cm$^{-3}$ in Perseus, and 390~cm$^{-3}$
in Ophiuchus.   To be identified as a core, therefore, individual
structures must have a mean density $n\gtrsim 2\times 10^4$~cm$^{-3}$,
and a contrast compared to the average background density of at least
30-100.  The mean cloud density is estimated from the total cloud mass
(\S~\ref{masssec}) and assumes a cloud volume of
$\mathrm{V}=\mathrm{A}^{1.5}$, where A is the area of the extinction
map within the $\av=2$ contour.  

Although we are primarily sensitive to cores with high density
contrast compared to the background, it is clear that there is
structure in the 1.1~mm map at lower contrasts as well, and that many
cores are embedded within lower density filaments.   The total mass in
each of the 1.1~mm maps, calculated from the sum of all pixels
$>5\sigma$, is approximately twice the mass in dense cores:
176\msun\ versus 92\msun\ in Serpens, 376\msun\ versus 278\msun\ in
Perseus, and 83\msun\ versus 44\msun\ in Ophiuchus, for ratios of
total 1.1~mm mass to total core mass of 1.9, 1.4, and 1.9
respectively.  Thus about half the mass detectable at 1.1~mm is not
contained in dense cores, but is rather in the ``foothills'' between
high density cores and the lower  density cloud medium.

Structures that meet the above sensitivity criteria and are identified
by our peak-finding routine  are considered cores.  Our peak-finding
method will cause extended filaments to be broken up into several
separate ``cores'' if there are local maxima in the filament separated
by more than one beam size, and if each has a well-defined centroid
(see \S~\ref{idsect}).   There is some question as to whether these
objects should be  considered separate sources or a single extended
structure, but we believe that our method is more reliable for these 
data than alternative methods such as Clumpfind \citep{will94}.  
In Paper~I we found that faint extended sources in our maps, which one 
would consider single if examining by eye, are often partitioned into 
multiple sources by Clumpfind.
Using our method, one filamentary
structure in  Serpens is broken up into several sources, as are two
filaments in Perseus.

Monte Carlo tests were done to quantify biases and systematic errors
introduced by the cleaning and iterative mapping process, which affect
the kind of structure we can detect.  Measured FWHM, axis ratios, and
position angles are not significantly affected by either cleaning or
iterative mapping for sources with FWHM$\lesssim 120''$; likewise, any
loss of flux for such sources has an amplitude less than that of the
rms noise.  Sources with FWHM$\gtrsim 200''$ are detected, but with
reduced flux density (by up to 50\%), and large errors in the measured
FWHM sizes of up to a factor of two. 

The limitation on measurable core sizes of approximately $120''$
corresponds to $3\times10^4$~AU in Perseus and Serpens, and
$1.5\times10^4$~AU  in Ophiuchus.   Note that these sizes are  of
order the median core separation in each cloud (see
\S~\ref{clustsect}), meaning  that we are just as likely to be limited
by the crowding of cores as by our sensitivity limits in the
measurement of large cores.  The dependence of measurable core size on
cloud distance  can be seen in Figure~\ref{mvscomp}, where the
completeness rises  steeply at smaller linear deconvolved sizes for
Ophiuchus than for Perseus or Serpens.  Thus we are biased against
measuring large cores in Ophiuchus compared to the other two clouds.
Although there are cores in our sample with sizes up to
$3\times10^4$~AU (Figure~\ref{mvscomp}),  most cores have sizes 
substantially smaller than the largest measurable value, again
indicating that we are not limited by systematics.

To summarize, the Bolocam 1.1~mm observations presented here naturally
pick out sub-structures in molecular clouds  with high volume density
($n\gtrsim 2\times 10^4$~cm$^{-3}$).  These millimeter cores have a
contrast of at least $30-100$ compared to the average cloud density as
measured by the visual  extinction map ($2-10\times10^3$~cm$^{-3}$).
Many cores are embedded in lower density extended structures,  which
contribute approximately half the mass measurable in the Bolocam
1.1~mm maps.  Finally, Monte Carlo tests indicate that we can detect
cores with intrinsic sizes up to approximately $120''$.  

\subsection{Distance effects}\label{convsect}

To test the effects of instrumental resolution and its dependence on
distance, we convolve the Ophiuchus map with a larger beam to simulate
putting it at approximately the same distance as Perseus and Serpens.
After convolving the unfiltered Ophiuchus map to $62''$ resolution, we
apply the optimal filter and re-compute the local rms noise, as
described for Serpens (\S~\ref{idsect}).  The pixel scale in the
convolved map is still  10\arcsec\ pixel$^{-1}$, but the resolution is
now $62''$ and the rms is lower than in the original map, with  a
median value of 17~mJy beam$^{-1}$ in the main L~1688 region. Source
detection and photometry is carried out in the same way as for the
original map, with the exception that a 62\arcsec\ beam is assumed.

We detect 26 sources in the degraded-resolution map, or 40\% fewer
than the 44 sources in the original map.  Therefore a number of
sources do become confused at lower resolution.  The basic source
properties for the original and degraded-resolution samples, including
angular deconvolved sizes,  axis ratios, mass distribution, and mean
densities, are compared in Figure~\ref{convplots}.  Here the angular
deconvolved size is defined as   $\theta_{dec} = \sqrt{
\theta_{meas}^2 -  B^2}$, where $\theta_{meas}$  is the geometric mean
of the measured minor and major axis FWHM sizes  and $B$ is the
pointing-smeared  beam FWHM ($32.5\arcsec$ and $62\arcsec$ for the
original and degraded resolution maps, respectively).

We note that the average source size in the degraded-resolution
map is nearly twice that in the original Ophiuchus map
(average angular deconvolved size of $61''$ versus $98''$).   Such a large
size difference cannot be fully accounted for by blending of sources,
as even isolated sources show the same effect.  This behavior provides
clues to the intrinsic intensity profile of the sources.  For example,
a gaussian intensity profile or a solid disk of constant intensity
will both have measured deconvolved sizes that are similar in maps
with $31''$ and $62''$ beams.  Conversely, a power law intensity
profile will have a larger measured size in the $62''$ resolution map.

The ratio of angular deconvolved size to beam size
($\theta_{dec}/\theta_{mb}$; Figure~\ref{sizetconv}) are similar for
the degraded-resolution (median $\theta_{dec}/\theta_{mb}=1.7$) and
original (median $\theta_{dec}/\theta_{mb}=1.5$) samples, further
evidence for power law intensity profiles.  An intrinsic gaussian or
solid disk intensity profile will result in $\theta_{dec}/\theta_{mb}$
values in the degraded-resolution map that are approximately half
those in the original map, while a $1/r^2$ intensity profile results
in similar values in the degraded-resolution and original map (0.9 
versus 1.3).  We discuss source profiles further in
Section~\ref{sizecompsect} below.

Sources in the degraded-resolution map appear slightly more elongated,
with an average axis ratio at the half-maximum contour of $1.3\pm0.2$
compared  to $1.2\pm0.2$ for the original map.   Larger axis ratios
are expected for blended sources in a lower-resolution map.  The slope
of the mass distribution is not significantly changed: $2.0\pm0.4$ for
the degraded-resolution sample compared to $2.1\pm0.3$ for the
original sample.  The factor of two increase in deconvolved sizes do
create lower mean densities in the degraded-resolution sample, however
(median density of $1.6\times10^5$~cm$^{-3}$ compared to
$5.8\times10^5$~cm$^{-3}$ for the original sample).  The effect of
resolution on mean densities will be discussed further in
\S~\ref{dencompsect}.

Given the small number of sources in the degraded-resolution map, we carry out
the three cloud comparison below using the original Ophiuchus map to
mitigate uncertainties from small number statistics.  Any notable
differences between the original and degraded-resolution Ophiuchus results will
be discussed where appropriate. 

\subsection{Physical Implications of Source Sizes and Shapes}\label{sizecompsect}

To compare sources in clouds at different distances, we first look at
the linear deconvolved size:  $D_{dec} = d \sqrt{ \theta_{meas}^2 -
  B^2}$, where $d$ is the cloud distance, $\theta_{meas}$ is measured
angular size, and $B$ is the beam size.  The distributions of
linear deconvolved source sizes for the three clouds are shown in
Figure~\ref{dsizecomp}, left.  Sources in Ophiuchus have smaller
deconvolved sizes than  those  in Perseus or Serpens by almost a
factor of two, with mean vales of $7.5\times10^3$~AU in Ophiuchus
compared to $1.2\times10^4$~AU in Serpens and $1.5\times10^4$~AU in
Perseus.   There is a systematic uncertainty in the deconvolved size
associated with the uncertainty in the effective beam size, which
becomes larger with larger pointing errors.  The effective beam in any
of the three clouds may be as large as $35\arcsec$, which would
decrease deconvolved sizes negligibly, by up to $10^3$~AU depending on the
distance and measured size.

While there are possible physical explanations for intrinsic size
differences, for instance a denser medium with a shorter Jeans length
should produce smaller cores on average, we are more likely seeing a
consequence of the higher linear  resolution in Ophiuchus, as
discussed in \S~\ref{convsect}.   Thus cores in Serpens and Perseus
would likely appear smaller if observed at higher resolution, and measured 
linear deconvolved sizes should be regarded as upper limits.   
To reduce the effects of distance, we examine the ratio of angular  
deconvolved size to beam size ($\theta_{dec}/\theta_{mb}$;
Figure~\ref{dsizecomp}, right).  We found in \S~\ref{convsect} that
$\theta_{dec}/\theta_{mb}$ does not depend strongly on the linear
resolution, but does depend on  the intrinsic source intensity
profile.

If the millimeter sources follow power law density distributions,
which do not have a well defined size, then \citet[][hereafter
  Y03]{young03}  show that $\theta_{dec}/\theta_{mb}$ depends on the
index of the power-law, and not on the distance of the source.  So,
for example, if sources in Perseus and Ophiuchus have the same
intrinsic power law  profile, the mean $\theta_{dec}/\theta_{mb}$
should be similar in the two clouds, and the mean linear deconvolved
size should be twice as small in Ophiuchus because it lies at half the
distance.   This is precisely the behavior we observe, suggesting that
many of the detected  1.1~mm sources have power law density profiles.

Considering that a number of the 1.1~mm sources have internal
luminosity sources (M. Enoch et al. 2007, in preparation), and that
protostellar envelopes are often well  described by power law profiles
\citep[][Y03]{shir02}, this is certainly a plausible scenario.
According to the correlation between $\theta_{dec}$ and density power
law exponent $p$ found by Y03,   median $\theta_{dec}/\theta_{mb}$
values of 1.7 in Perseus, 1.5 in Ophiuchus and 1.3 in Serpens would
imply average indices of $p=1.4$, 1.5, and 1.6 respectively.   These
numbers are  consistent with mean $p$ values found from radiative
transfer modeling of  Class 0 and Class I envelopes \citep[$p\sim
  1.6$,][Y03]{shir02},  although the median for those samples is
somewhat higher ($p\sim 1.8$).  Note that source profiles could
deviate from a power-law on scales much smaller than the beam size, or
on scales larger than our size sensitivity ($200''$)  without
affecting our conclusions.

Perseus displays the widest dispersion of angular sizes, ranging
continuously from $1-4 \theta_{mb}$.  By contrast, more than half the
sources in Serpens and Ophiuchus are  within $0.5 \theta_{mb}$ of
their respective mean values.  Although there is a group of Ophiuchus
sources  at large sizes in Figure~\ref{dsizecomp}, note that the
degraded-resolution Ophiuchus sample  displays a very narrow range of
sizes (Figure~\ref{sizetconv}), similar to Serpens.  The observed size
range in Perseus would correspond to a wide range of power law
indices, from very shallow ($p\sim1$) to that of a singular isothermal
sphere ($p=2$).  A more likely possibility, however, is that sources
with large $\theta_{dec}/\theta_{mb}$ do not  follow power law density
profiles. 

The axis ratio at the half-maximum contour  is a simple measure of
source shape.  Figure~\ref{axiscomp} shows the distribution of 1.1~mm
source axis ratios in the three clouds.  Our simulations suggest that
axis ratios of up to 1.2 can be introduced by the data reduction
(Paper~I),  so we consider sources with an axis ratio $<1.2$ to be
round, and  those with a ratio $>1.2$ to be elongated.  Sources in
Ophiuchus  tend to be round, with a mean axis ratio of 1.2, but note
that the mean axis ratio in the degraded-resolution Ophiuchus sample
is 1.3.   The average axis ratio in Serpens  is 1.3, and Perseus
sources exhibit the largest axis ratios with a mean  of 1.4 and a tail
out to 2.7.  

We found  in Paper~II that Ophiuchus sources were more elongated at
the $4\sigma$ contour than at the half-max contour, as would be the
case for round cores embedded in more elongated filaments.  A similar
situation is seen in Serpens; the average axis ratio at the $4\sigma$
contour is 1.4 in all three clouds.  Thus cores in Perseus are
somewhat elongated on average, while objects in Serpens and Ophiuchus
appear more round at the half-max contour but elongated at the
$4\sigma$ contour, suggesting round cores embedded in filamentary
structures.

In addition to angular sizes, Y03  also note a relationship between
axis ratio and  density power law exponent, finding that aspherical
sources are best modeled with shallower density profiles.  The inverse
proportionality between $p$ and axis ratio demonstrated in  Figure~25
of Y03 suggests power law indices in all three clouds between 1.5 and
1.7.   These values are consistent with those inferred from the
average angular deconvolved source sizes, and the wider variation of
axis ratios in Perseus  again points to a larger range in $p$ for that
cloud.

\subsection{Densities and the Mass versus Size Distribution}\label{dencompsect}

Mean densities calculated using linear deconvolved FWHM sizes  appear
to be significantly higher in Ophiuchus, where the median of the mean
densities of the sample is $5.4\times10^5$~cm$^{-3}$, than in Serpens
(median density $2.2\times10^5$~cm$^{-3}$) or Perseus  (median density
$1.6\times10^5$~cm$^{-3}$), as seen in Figure~\ref{dencomp}, left.
There is a large scatter with  standard deviation of order twice the
mean value in all three clouds.  Sources in Ophiuchus tend to be less
massive than in the other two clouds, so the larger mean densities can
be entirely attributed to smaller deconvolved sizes in  the Ophiuchus
sample, which are sensitive to the  shape of the intrinsic density
distribution (Y03).  As noted in \S~\ref{convsect}, linear resolution
has a strong systematic effect on deconvolved sizes, and consequently
on mean densities.  The median density of the degraded-resolution Ophiuchus
sample is $1.6\times10^5$~cm$^{-3}$,  similar to both Perseus and
Serpens.

We additionally calculate mean densities using the full-width at
$4\sigma$ size  rather than the FWHM size (Figure~\ref{dencomp},
right)  to test the hypothesis that  mean density differences are
largely an effect of how source sizes are measured.  Using this
definition, differences between the clouds are less  pronounced, with
median densities of $1.2 \times 10^5$~cm$^{-3}$ in Perseus,  $1.3
\times 10^5$~cm$^{-3}$ in Serpens, and $2.0 \times 10^5$~cm$^{-3}$ in
Ophiuchus.   These numbers suggest that source mean densities are less
dependent  on cloud distance when measured at the radius where the
source merges into the  background, rather than at the half-max. 

Figure~\ref{mvscomp2} again displays the source total mass versus size
distribution, using the angular deconvolved size ($\theta_{dec}$, in
units of the beam size) rather than the linear deconvolved size.
With the exception of two low mass compact sources, 1.1~mm sources in
Serpens exhibit a wide range of masses and a narrow range of sizes.
Sources in Perseus, in contrast, demonstrate a wide range in both mass
and size.  This difference likely reflects a wider variety of physical
conditions in the Perseus cloud, including the existence of more
sources outside the main cluster regions.  In such lower density
regions, sources may be more extended than in densely populated
groups.  This idea is supported by our observations: Perseus sources
within the NGC~1333 region are smaller on average (mean size of
$59\pm13''$) than those in the rest of the cloud (mean size of
$71\pm21''$).  Sources in Ophiuchus have a more bimodal distribution,
with the majority  occupying a narrow range of sizes, and a smaller
group with  $\theta_{dec}>3\theta_{mb}$.

\subsection{Fragmentation and the Core Mass Distribution}\label{masscompsect}

Differential ($dN/dM$) core mass distributions (CMDs) of 1.1~mm
sources in Serpens, Perseus, and Ophiuchus are shown in
Figure~\ref{masscomp}, with those of Perseus and Ophiuchus  scaled
down for clarity.  Error bars reflect $1\sigma$ $\sqrt{N}$ statistical
uncertainties.  Dashed lines indicate empirical 50\% completeness
limits for average sized sources in each cloud, which are determined
as described in \S~\ref{coresect} and in Paper~I.  Mass distributions
include all 1.1~mm cores in each cloud, including those that may be
associated with embedded protostellar sources.  

The shape of the Ophiuchus and Perseus mass distributions are quite
similar above their respective completeness limits ($M \gtrsim 0.5
\msun$ in Ophiuchus and $M \gtrsim 0.8 \msun$ in Perseus).   Fitting a
power law ($dN/dM \propto M^{-\alpha}$) to the CMDs, we find that they
both have a best fit slope of $\alpha=2.1$, although the error is
larger on the slope for Ophiuchus ($\sigma_{\alpha} = 0.3$) than for
Perseus ($\sigma_{\alpha} = 0.1$).  The slope of the Serpens CMD is
marginally different ($\alpha = 1.6 \pm 0.2$), being flatter than in
the other two clouds by approximately  $2 \sigma$.

\begin{figure}
\plotone{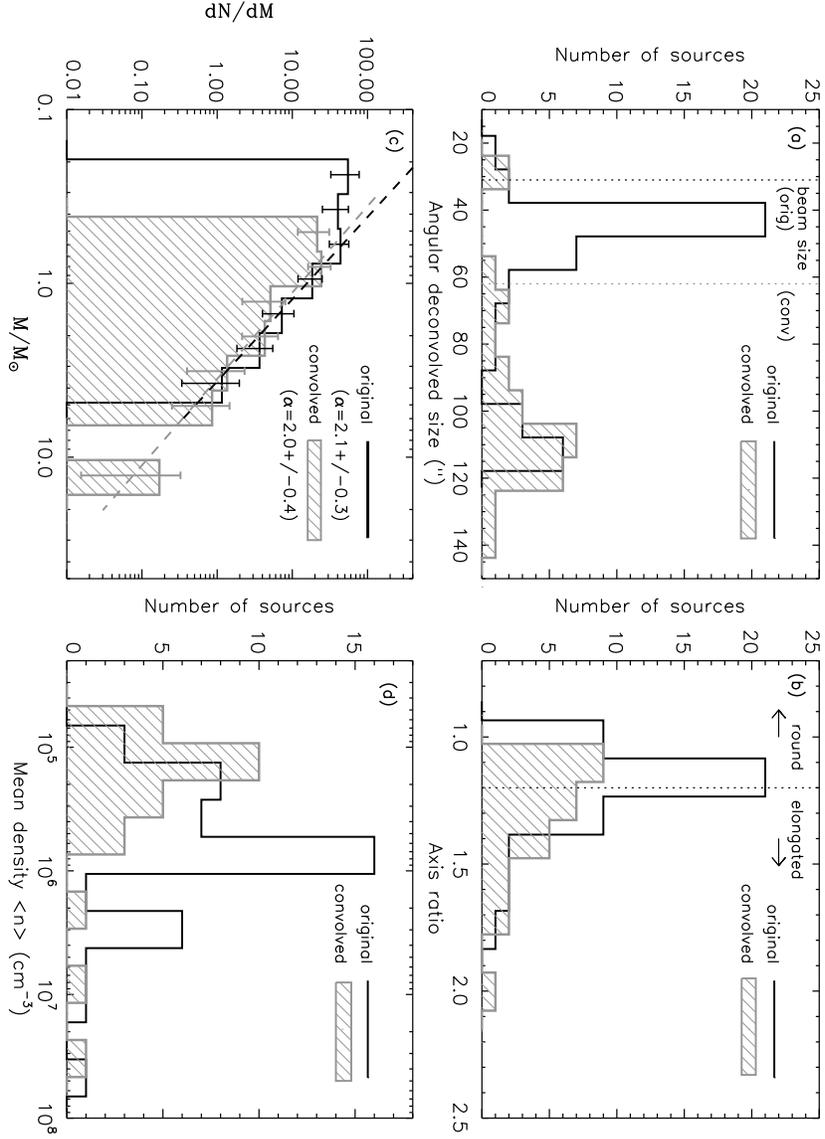}
\caption{Comparison of the basic source properties for the original
  (black curve) and degraded-resolution, or convolved, (gray hatched
  curve) Ophiuchus source samples, which contain 44 and 26 cores
  respectively. (\textit{a}) Measured angular deconvolved sizes are
  larger in the degraded-resolution map than in the original map by
  approximately a factor of two (respective mean values of $98''$ and
  $61''$). (\textit{b}) Sources tend to be slightly more elongated in
  the degraded-resolution map, with an average axis ratio of
  $1.3\pm0.2$ compared to  $1.2\pm0.2$ in the original
  map. (\textit{c}) The slope of the CMD is not significantly changed
  for the degraded-resolution sample, but a number of low mass cores
  are blended into a few higher mass sources.  (\textit{d})  Larger
  deconvolved sizes lead to lower mean densities for the
  degraded-resolution sample (median values of
  $1.6\times10^5$~cm$^{-3}$ and
  $5.8\times10^5$~cm$^{-3}$).}\label{convplots}
\end{figure}

\begin{figure}
\plotone{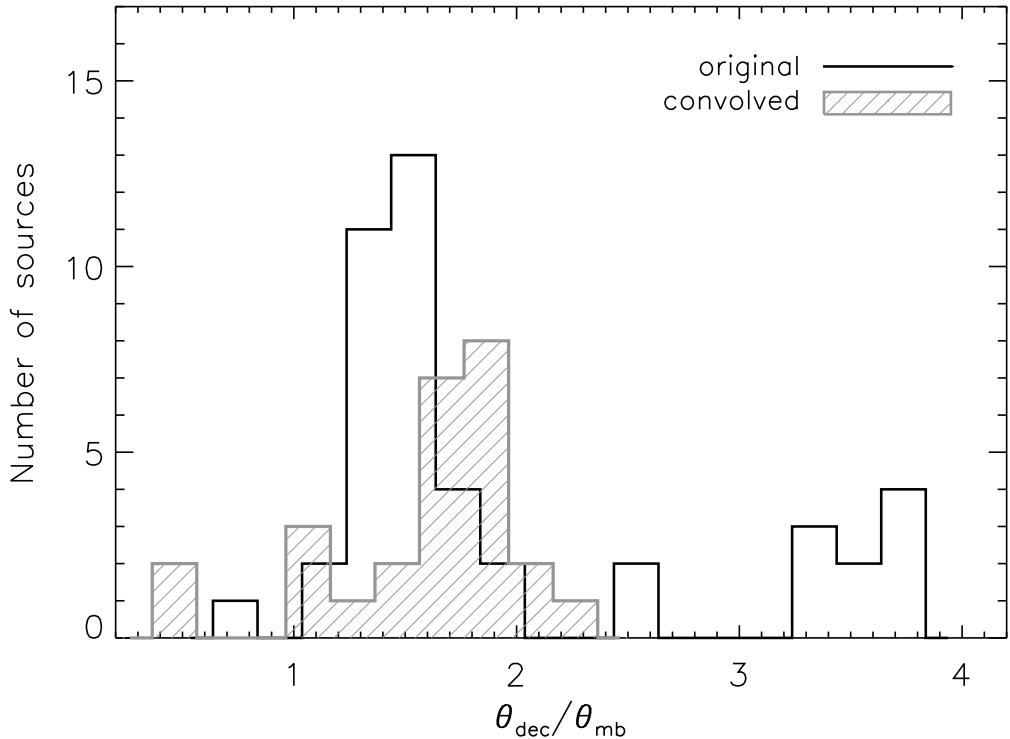}
\caption{Ratio of angular deconvolved size to beam size
  ($\theta_{dec}/\theta_{mb}$) for the original and degraded-resolution,
  or convolved, Ophiuchus maps.  Note that $\theta_{mb}$ is $31''$ for
  the original map  and $62''$ for the degraded-resolution  map.
  Measured  $\theta_{dec}/\theta_{mb}$ values are similar for the
  degraded-resolution (median $\theta_{dec}/\theta_{mb}=1.7$) and
  original (median $\theta_{dec}/\theta_{mb}=1.5$) samples, providing
  evidence for power law intensity profiles.  An intrinsic gaussian or
  solid disk intensity profile will result in
  $\theta_{dec}/\theta_{mb}$ values in the degraded-resolution map
  that are nearly half those in the original map, while a $1/r^2$
  intensity profile results in similar values in the
  degraded-resolution and original map (0.9 vs
  1.3).}\label{sizetconv}
\end{figure}

\begin{figure}
\plottwo{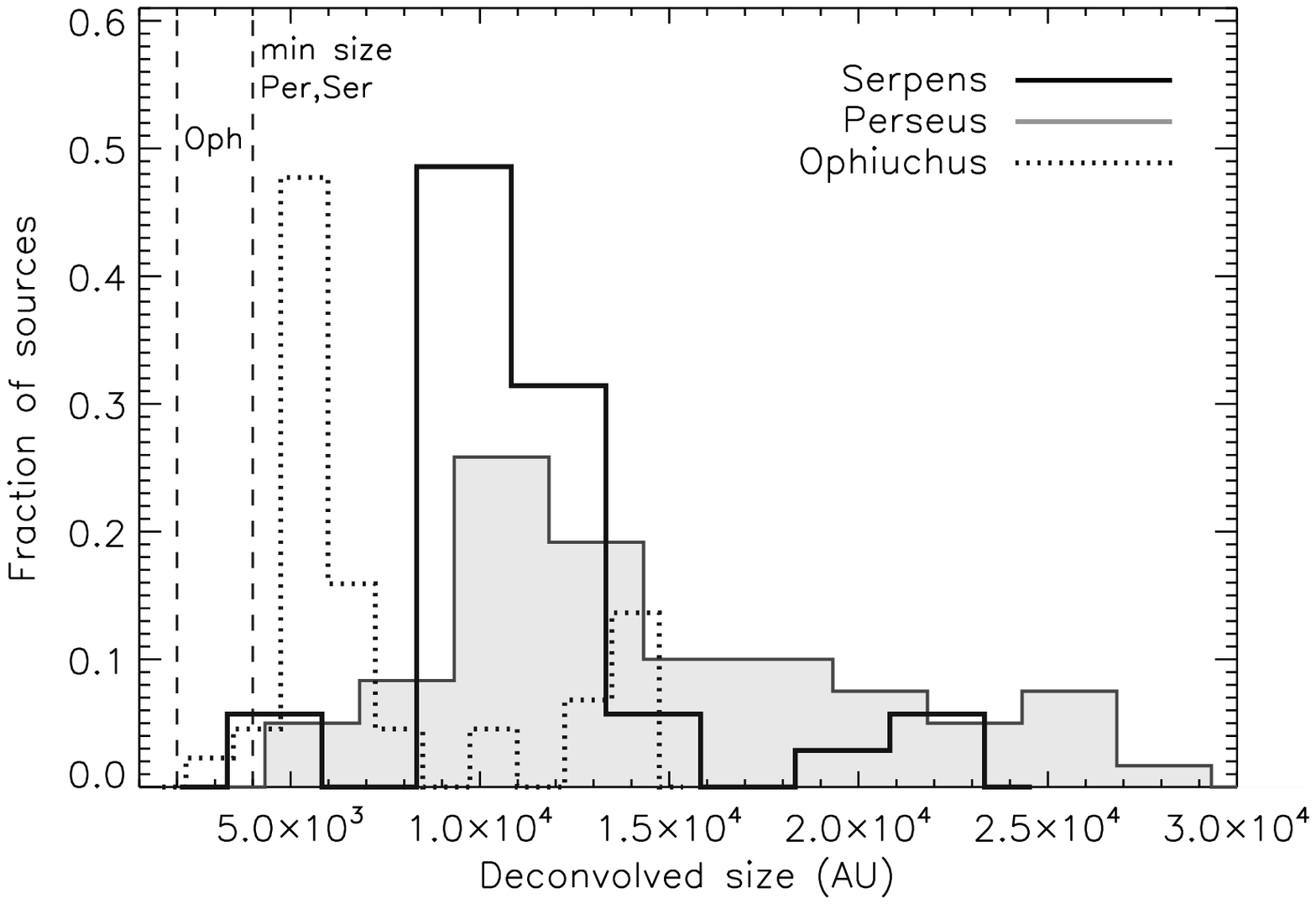}{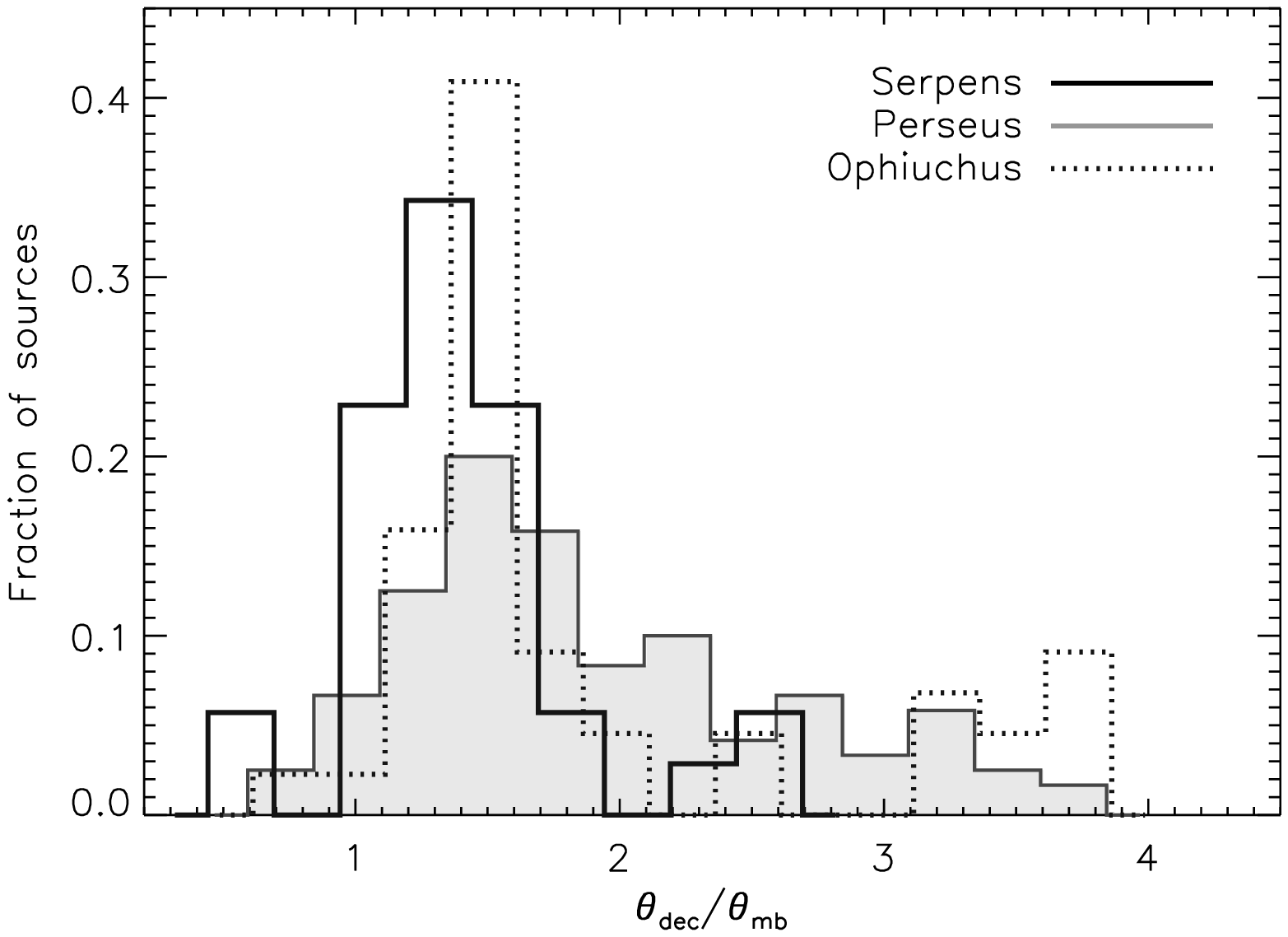}
\caption{\textit{Left}: Comparison of the distribution of deconvolved
  linear source sizes in Serpens (solid line), Perseus (shaded), and
  Ophiuchus (dotted line).  Histograms are plotted as the fraction of
  total sources in the cloud as a function of deconvolved size in AU.
  Estimates of the minimum resolvable source size (dashed lines) are
  based on smearing of the beam by pointing errors.  Although measured
  angular source sizes in Ophiuchus are similar to those in Perseus and
  Serpens, the deconvolved sizes are much smaller due to the different
  cloud distances.  \textit{Right}: Similar, but sizes are measured as
  the angular deconvolved size in units of the beam FWHM.  For isolated 
  sources with power law density profiles,  $\theta_{dec}/\theta_{mb}$ is
  inversely proportional to the power law index and  is independent of
  cloud distance \citep{young03}.
\label{dsizecomp}}
\end{figure}

\begin{figure}
\plotone{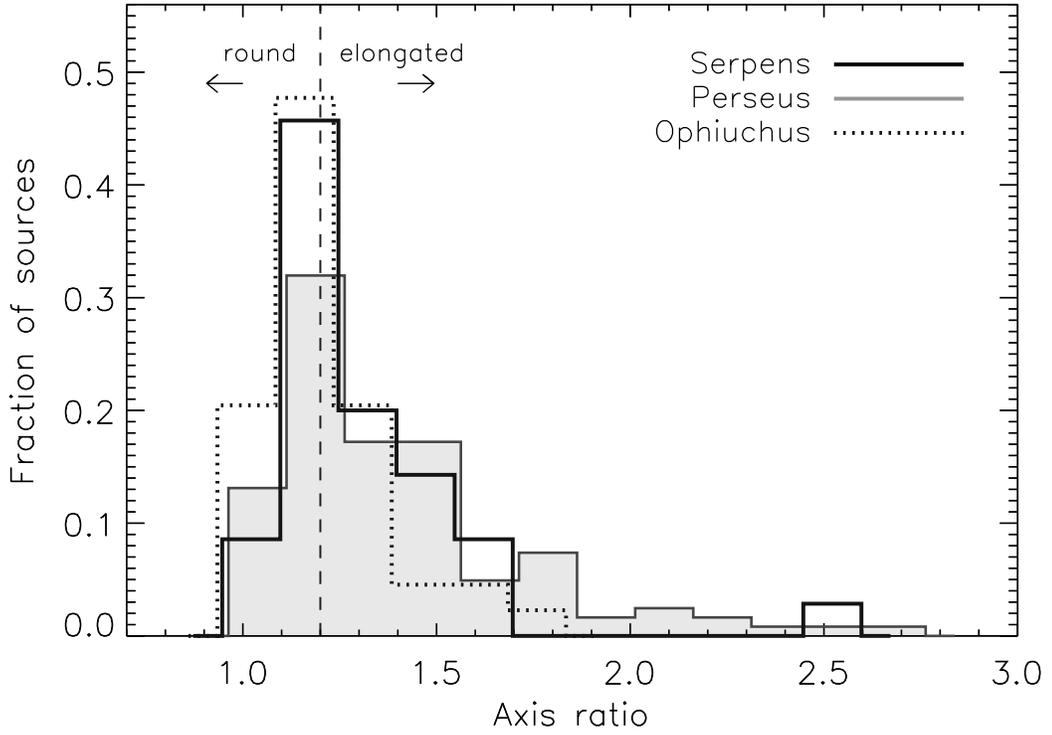}
\caption{Comparison of the distribution of axis ratios, where the
  ratio is calculated at the half-max contour.   Axis ratios $<1.2$
  are considered round, and $>1/2$ elongated, based  on Monte Carlo
  simulations.  Sources are primarily round in Ophiuchus and Serpens,
  with mean axis  ratio of 1.2 and 1.3, respectively.  Perseus
  exhibits the most  elongated sources, with a mean axis ratio of 1.4
  and a distribution tail extending up to 2.7.  \label{axiscomp}}
\end{figure}

\begin{figure}
\plottwo{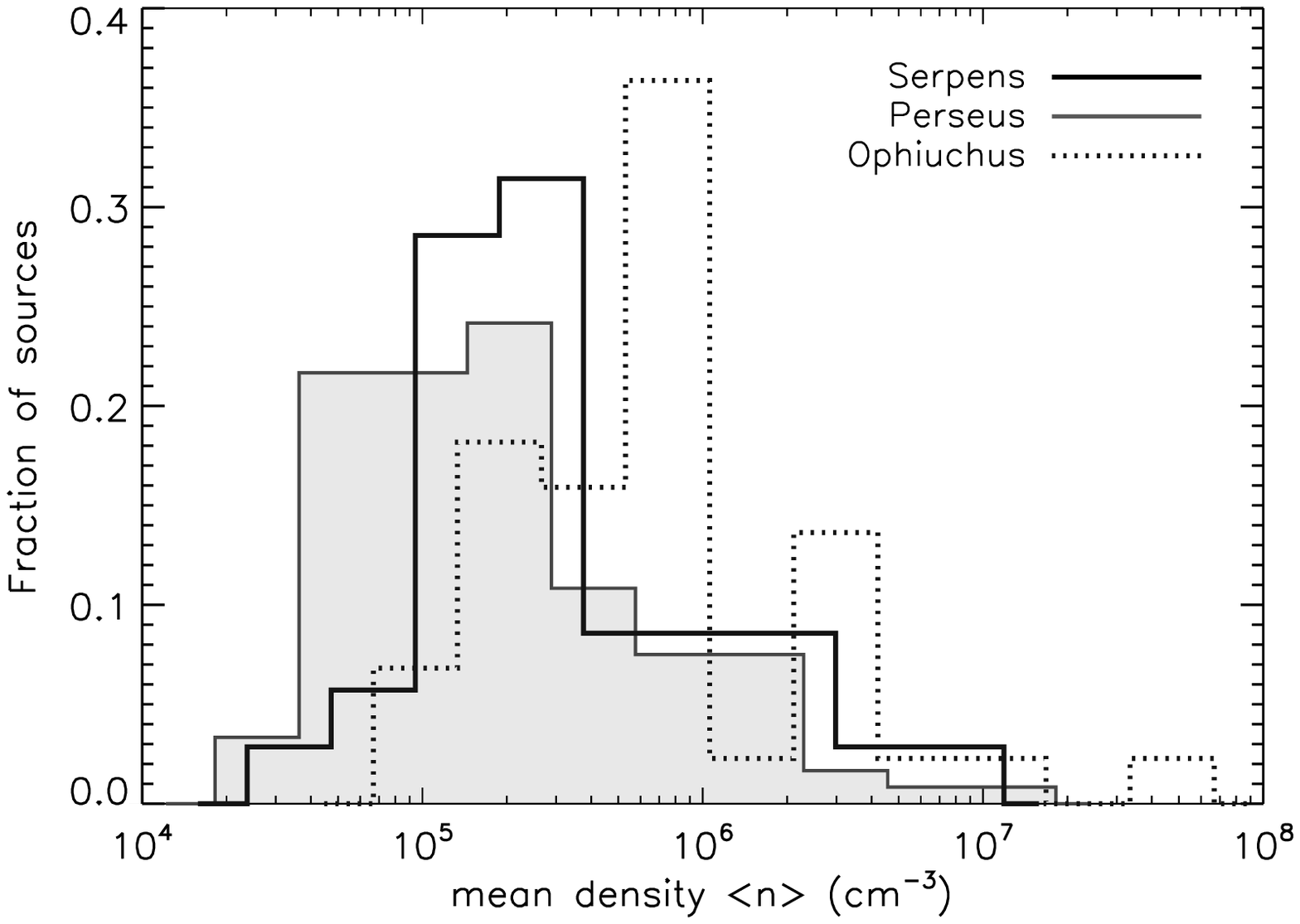}{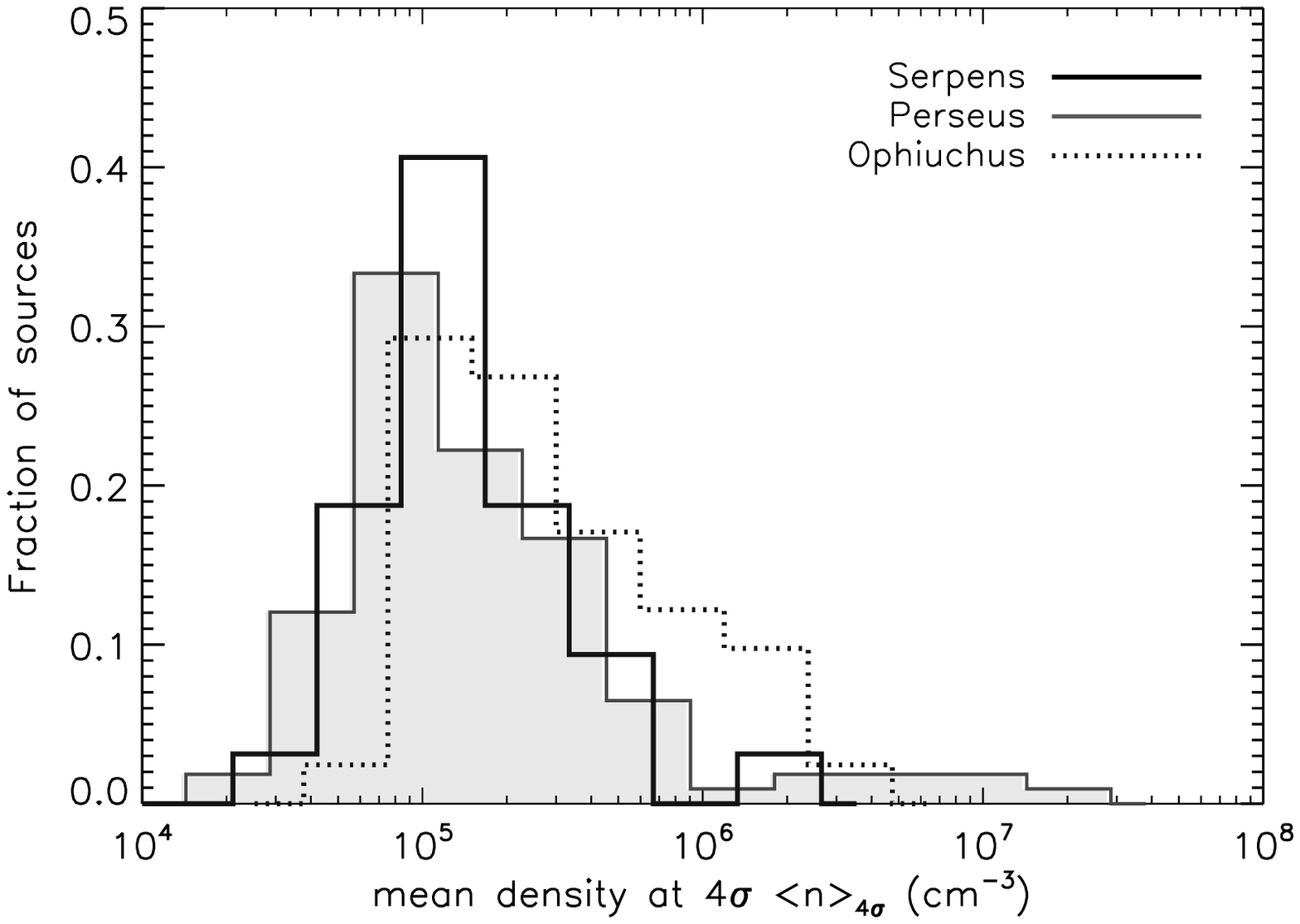}
\caption{Comparison of the distributions of core mean densities.
  \textit{Left:}  The mean density is calculated from the total mass
  (Table~\ref{phottab}) and the linear deconvolved source size,
  assuming spherical cores.  Despite the lower masses of cores in
  Ophiuchus compared to Serpens and Perseus, mean densities in
  Ophiuchus are higher on average due to small linear deconvolved
  sizes (see Figure~\ref{dsizecomp}).   The mean densities in the
  degraded-resolution Ophiuchus sample are more consistent with Perseus and
  Serpens, suggesting cloud distance has a strong effect.
  \textit{Right:} Similar, except that the mean density is calculated
  using the full-width at $4\sigma$ size rather than of FWHM size.
  Differences between the clouds are diminished; thus the mean density
  is less dependent on cloud distance when the source size is measured
  at  the radius where the source merges into the background, rather
  than at  the half-max.
\label{dencomp}}
\end{figure}

\begin{figure}
\plotone{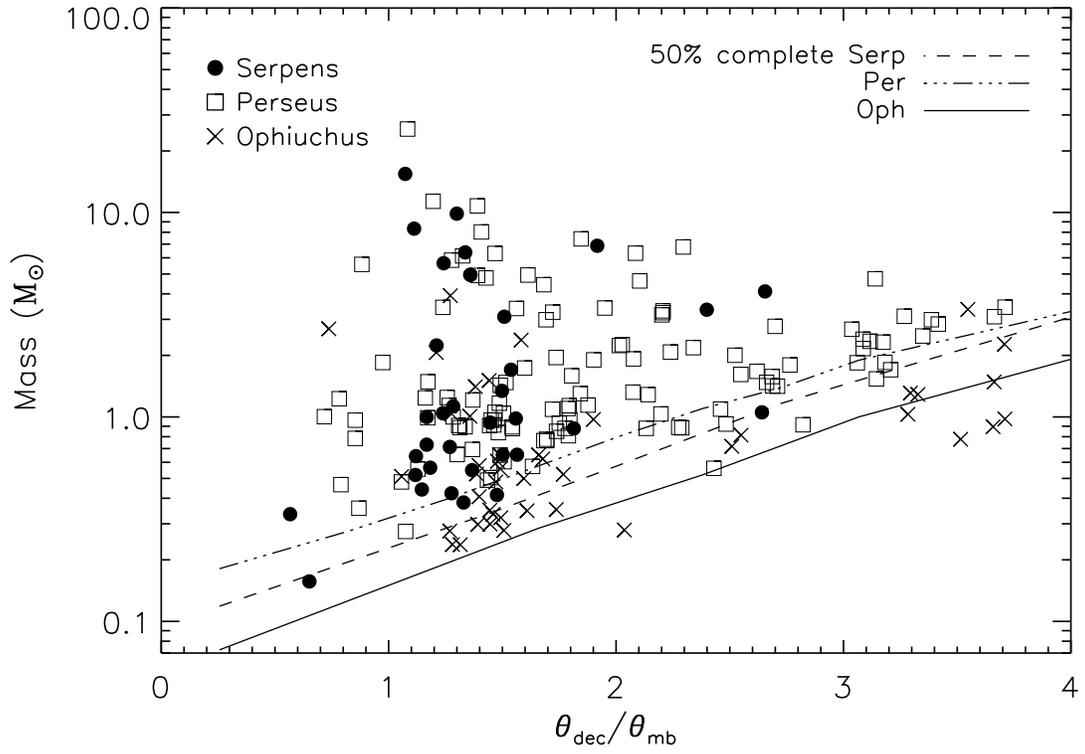}
\caption{ Source total mass versus angular deconvolved size (in units
  of  the beam FWHM) for Serpens, Perseus, and Ophiuchus.  The angular
  size is less dependent on cloud distance than the linear deconvolved
  source size (\S~\ref{convsect}).  
  Sources in Perseus fill more of the parameter space than those in 
  Serpens or Ophiuchus, suggesting a wider range of physical conditions 
  in that cloud.
  Empirically derived 50\% completeness limits are also shown (lines).
\label{mvscomp2}}
\end{figure}

\begin{figure}
\plotone{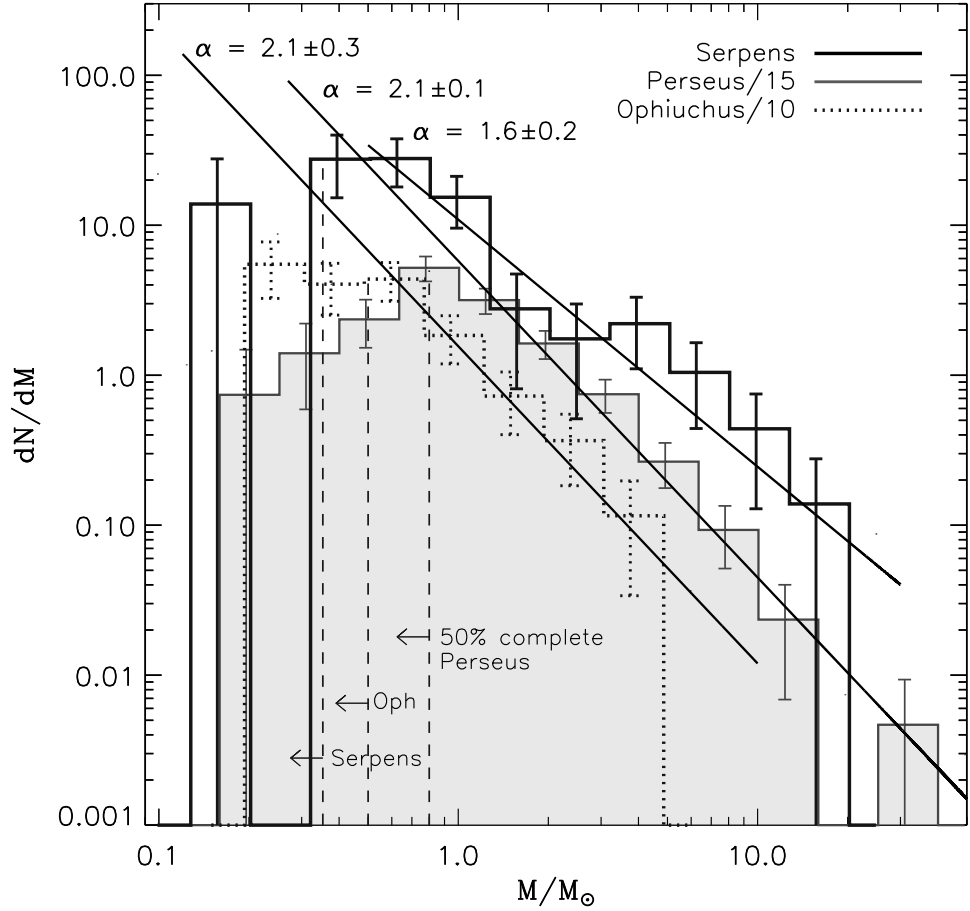}
\caption{ Comparison of the differential core mass distributions
  (CMDs) of sources in Serpens,  Perseus, and Ophiuchus.  The
  Ophiuchus curve has been scaled by 1/10, and the Perseus curve by
  1/15 for clarity.  Uncertainties reflect $\sqrt{N}$ counting
  statistics only.  Vertical dotted lines show the 50\% completeness
  limits for average sized sources in each cloud  (55\arcsec\ FWHM
  in Serpens, 68\arcsec\ in Perseus and 69\arcsec\ in Ophiuchus).  The
  shape of the Perseus and Ophiuchus distributions are quite similar 
  ($\alpha = 2.1$) in the region where both CMDs are complete.  The
  shape of the Serpens CMD is marginally different
  ($\alpha=1.6$) from the other two clouds. \label{masscomp}  }
\end{figure}

The two-sided Kolmogorov-Smirnov test indicates a high probability
(46\%) that the Perseus and Ophiuchus mass distributions are
representative  of the same parent population.  Conversely, the
probabilities that the Serpens core masses are sampled from the  same
population as the Perseus (prob$=12$\%) or Ophiuchus (prob$=5$\%)
masses  are much lower.  Although it is possible that the lower linear
resolution in Perseus and Serpens has lead to larger masses via
blending, the test we conducted to increase the effective beam size in
the Ophiuchus map  did not appreciably change the shape of the
Ophiuchus mass distribution (see Figure~\ref{convplots}). 

If the shape of the CMD is a result of the fragmentation process, then
the slope of the CMD can be compared to models, e.g. of turbulent
fragmentation.  \citet{padoan02} argue that turbulent fragmentation
naturally produces a power law with $\alpha = 2.3$, consistent with
the slopes we measure in Perseus and Ophiuchus ($\alpha=2.1$), but not
with Serpens ($\alpha=1.6$).  Recently \citet[][hereafter
  BP06]{paredes06}  have questioned that result, finding that the
shape of the CMD depends strongly on the Mach number of the turbulence
in their simulations.  The BP06 smoothed particle hydrodynamics (SPH) 
simulations show  that higher Mach
numbers result in a larger number of sources with lower mass and a
steep slope at the high mass end (their Figure~5).   Conversely, lower
Mach numbers favor sources with higher mass, resulting in a smaller
number of low mass sources, more high mass cores, and a shallower
slope at the high mass end.  

Using an analytic argument, \citet{padoan02} also note a relationship
between core masses and Mach numbers, predicting that  the mass of the
largest core formed by turbulent fragmentation should be inversely
proportional to the square of the Alfv\'{e}nic Mach  number  on the
largest turbulent scale  $\mathcal{M}_{A}^2$.  Given that our ability
to accurately measure the maximum core mass is  limited by resolution,
small number statistics, and cloud distance  differences, we focus
here on the overall CMD shapes.

To compare our observational results to the simulations of BP06,  we
estimate the sonic Mach number $\mathcal{M} = \sigma_v/c_s$ in each
cloud.  Here $\sigma_v$ is the observed rms velocity dispersion, $c_s
= \sqrt{kT/\mu m_H}$ is the isothermal sound speed, and $\mu = 2.33$
is the mean molecular weight per particle.   Large $^{13}$CO maps of
Perseus and Ophiuchus observed with FCRAO at a resolution of
44\arcsec\ are publicly available as part of ``The COMPLETE Survey of
Star Forming Regions''\footnote{http://cfa-www.harvard.edu/COMPLETE/}
\citep[COMPLETE;][]{good04,ridge06}.  Average observed rms velocity
dispersions kindly provided by the COMPLETE team are $\sigma_v =
0.68$~km~s$^{-1}$ in Perseus, $\sigma_v = 0.44$~km~s$^{-1}$ in
Ophiuchus, and  $\sigma_v = 0.92$~km~s$^{-1}$ in Serpens (J. Pineda,
personal communication).  These were measured by masking out all
positions in the map that have peak temperatures with a S/N less than
10, fitting a Gaussian profile to each, and taking an average of the
standard deviations.

We note that the value of $\sigma_v= 0.68$~km~s$^{-1}$ for Perseus is
smaller than a previous measurement of $\sigma_v= 2.0$~km~s$^{-1}$
based on AT\&T Bell Laboratory 7~m observations of a similar area of
the cloud \citep{pad03}.   The smaller value derived by the COMPLETE
team is most likely a  consequence of the method used: a linewidth is
calculated at every  position and then an average of these values is
taken.  In contrast to calculating the width of the averaged spectrum
this method removes the effects of velocity gradients across the
cloud.  The different resolutions of the surveys (44\arcsec\ and 0.07
km~s$^{-1}$ for the COMPLETE observations, 100\arcsec\ and 0.27
km~s$^{-1}$ for the Padoan et al. observations) may also play a role.
Sources of uncertainty in the linewidth measurement are not
insignificant, and include the possibility that $^{13}$CO is optically
thick to an unknown  degree, and the fact that line profiles are not
necessarily well  fit by a gaussian, especially in Perseus where the
lines sometimes  appear double-peaked.  

Assuming that the sound speed is similar in all three clouds, the
relative velocity dispersions suggest that turbulence is more
important in Serpens than in Perseus or Ophiuchus, by factors of
approximately 1.5 and 2,  respectively.  Mach numbers calculated using
the observed $\sigma_v$ and assuming a gas kinetic temperature of 10~K
are $\mathcal{M}=4.9$ (Serpens), $\mathcal{M}=3.6$ (Perseus), and
$\mathcal{M}=2.3$ (Ophiuchus).  We focus on Serpens and Ophiuchus, as
they are the most different.  Serpens is observed to have a higher
Mach number than  Ophiuchus, but the CMD indicates a larger number of
high mass cores,  a shallower slope at the high mass end, and a dearth
of low mass  cores compared to Ophiuchus.  This result is contrary to
the trends found by BP06, which would predict a steeper slope and more
low mass cores in Serpens.\footnote{It should be noted that the shape of the CMDs presented by BP06 depend on the type of numerical code used, and not on the cloud Mach number alone.  Here we have compared to the BP06 SPH simulations (their Figure~5); for the BP06 total variation diminishing (TVD) method (their Figure~4), a larger number of low mass cores are again seen for higher Mach numbers, but the CMD slopes at the high mass end are independent of Mach number.}

One core of relatively high mass is measured in the degraded-resolution 
map of Ophiuchus, as is expected for blending at  lower
resolution (Figure~\ref{convplots}), making the difference between the
Serpens and Ophiuchus CMDs less dramatic.  The degraded-resolution CMD
for Ophiuchus still has a steeper slope ($\alpha=2.0\pm0.4$) than  the
Serpens CMD, however, and a larger fraction of low mass sources: 81\%
of sources in the degraded-resolution Ophiuchus sample have $M<2M_{\sun}$
compared to 49\% in the Serpens sample.

More accurate measurements of the Mach number and higher resolution
studies of the relative CMD shapes will be necessary to fully test
the BP06  prediction, given that uncertainties are  currently too
large to draw firm conclusions.  As both  numerical simulations and
observations improve, however, the observed CMD and  measurements of
the Mach number will provide a powerful constraint on turbulent star
formation simulations. 

Comparing the shape of the CMD to the stellar initial mass function
(IMF) may give insight into what determines final stellar masses:  the
initial fragmentation  into cores, competitive accretion, or feedback
processes.   The shape of the local IMF is still uncertain
\citep{scalo05}, but  recent work has found evidence for a slope of
$\alpha=2.5-2.8$ for  stellar masses $M \gtrsim 1 \msun$, somewhat
steeper than the slopes we observe in all three clouds.  For example,
\citet{reid02} find $\alpha = 2.5$ above 0.6 \msun, and $\alpha = 2.8$
above 1 \msun. \citet{chab03} suggests $\alpha = 2.7$
($M>1~M_{\sun}$), while \citet{schroder03} finds $\alpha = 2.7$ for
$1.1 < M < 1.6$ \msun\ and $\alpha = 3.1$ for $1.6 < M < 4$ \msun.
For reference, the Salpeter IMF has a slope of $\alpha=2.35$
\citep{sal55}.

For comparison to the IMF, we would ideally like to construct a CMD
that  includes starless cores only, so that it is a measure of the
mass  initially available to form a star.  Although we cannot separate
prestellar cores from more evolved objects  with millimeter data
alone, we are currently comparing the Bolocam maps with c2d
\textit{Spitzer} Legacy maps of the same regions, which will allow us
to distinguish prestellar cores from those with internal luminosity
sources  (M. Enoch et al. 2007, in preparation).

\subsection{Clustering}\label{clustsect}

We use the two-point correlation function, 
\begin{equation}
w(r) = \frac{H_s(r)}{H_r(r)} - 1,
\end{equation}
as a quantitative measure of the clustering of cores.   Here $H_s(r)$
is the number of core pairs with separation between $log(r)$ and
$log(r)+dlog(r)$, and $H_r(r)$ is similar but for a random
distribution (see Paper~I).  Thus $w(r)$ is a measure of excess
clustering over a random distribution, as a function of separation.

\begin{figure}
\plotone{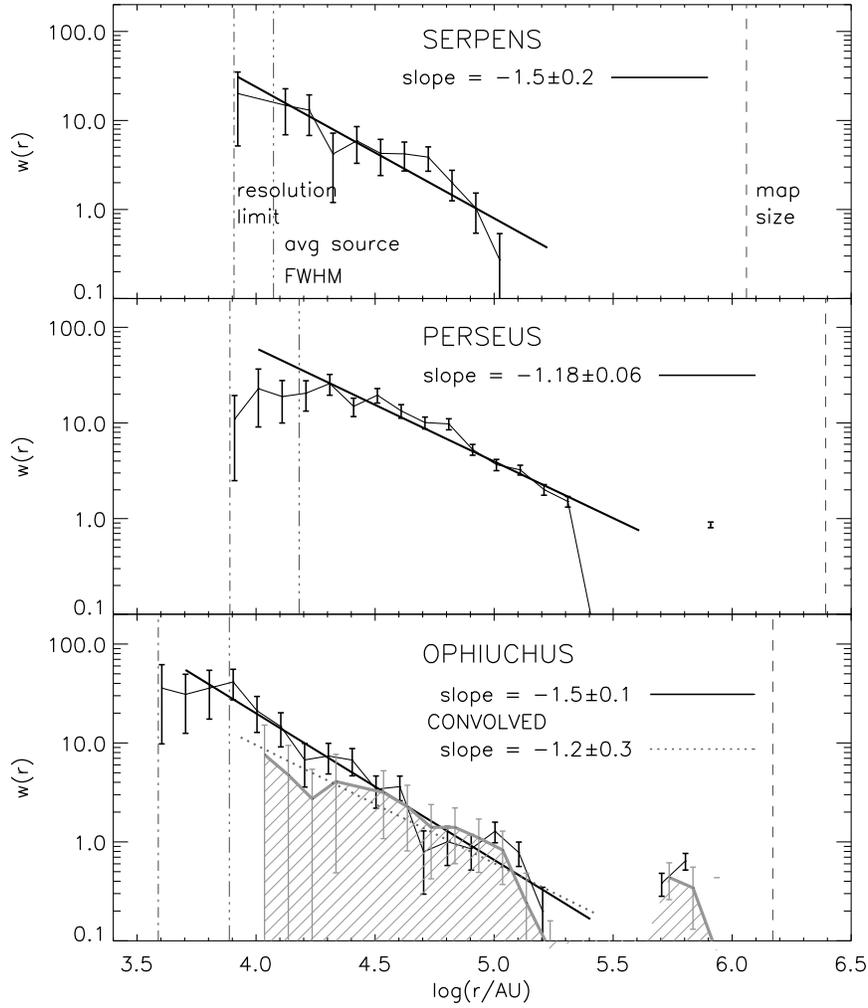}
\caption{ Comparison of the correlation function $w(r)$ for the three
  clouds, with power law fits.  For each cloud the average source FWHM
  size, resolution limit, and linear map size in AU are indicated.
  The best fit slope to  $w(r) \propto r^n$ is $n=-1.5\pm 0.2$ in
  Serpens, $n=-1.18\pm 0.06$ in  Perseus, and $n=-1.5\pm 0.1$ in
  Ophiuchus.   The slope found by \citet{john00} for a smaller map of
  Ophiuchus is shallower ($-0.75$).  The slope of $w(r)$ derived for
  the degraded-resolution Ophiuchus sample (hatched gray curve) is
  also shallower ($-1.2\pm0.3$), but the curve appears consistent
  with the original sample.  Differences in the slope of the
  correlation function likely trace differences in clustering
  properties of millimeter cores in the clouds.  \label{corrncomp}}
\end{figure}

Figure~\ref{corrncomp} plots the cloud two-point correlation
functions, with the best fit power laws, $w(r) \propto r^n$, shown.
The average linear source FWHM size is indicated, as well as the
minimum possible separation (the beam size), and the linear map  size.
The best fit slopes to $w(r)$ for Serpens ($n=-1.5 \pm 0.2$), Perseus
($n=-1.18 \pm 0.06$), and Ophiuchus ($n=-1.5 \pm 0.1$) are consistent
within $2\sigma$, but there is an indication both from the slopes and
a visual examination of  the plot that $w(r)$ falls off more steeply
in Serpens and Ophiuchus than in Perseus.   A shallower slope suggests
that clustering remains strong out to larger scales in Perseus than in
the other two clouds.

We do find that a broken power law with slopes of $n=-0.75$
($1.3\times 10^4 < r < 6.3\times 10^4$AU) and $n=-3.3$  ($r >
6.3\times 10^4$AU) is a better fit to the Serpens  correlation
function than a single power law,  indicating that  clustering remains
strong to  intermediate scales, as in Perseus, but then drops off
quickly.  Also shown in Figure~\ref{corrncomp} is $w(r)$ for the
degraded-resolution  Ophiuchus sample (gray hatched curve).   The
degraded-resolution sample is  fit by a shallower slope
($n=-1.2\pm0.3$) because there are fewer  sources at small
separations, but it is broadly consistent with the original Ophiuchus
sample.

Our general conclusion that  clustering of 1.1~mm sources remains
strong over larger scales in Perseus than in Serpens and Ophiuchus can
also be reached by visually examining each map. Perseus has highly
clustered regions spread over a larger area as well as a number of
more distributed sources, whereas Serpens and Ophiuchus have a single
main cluster with fewer small groups spread throughout the cloud.

A few caveats should be noted here.  First, the fact that clustering
seems to extend over larger scales in Perseus could simply be due to
the fact that Perseus has more widely separated regions of star
formation, and the physical association of all these regions has not
been firmly established.  Second, the map of Serpens covers a much
smaller linear area than that of Perseus (30~pc$^2$ versus
140~pc$^2$);  we need to be sure that the steepening of the slope in
Serpens is not caused by the map size.   This was confirmed by taking
a piece of the Perseus map  equal in size to the Serpens map and
recalculating $w(r)$.  Although the  value of the slope  changed
slightly, the smaller map size did not cause the slope to steepen at
large separations.  A more serious issue is that the overall
amplitude, but not the slope,  of $w(r)$ depends on  how large an area
the random distribution $H_r(r)$ covers.  We choose each random
distribution such that the largest pair separation is similar  to the
largest pair separation in the real data.  For Serpens and Ophiuchus
this means that the random distribution does not cover the entire
observed area.

Finally, we look at two other measures of clustering:  the peak number
of cores per square parsec, and the median separation of cores.
Although the median separation of cores is much smaller in Ophiuchus
($6.2\times10^3$~AU) than in Perseus ($2.6\times10^4$~AU) or Serpens
($2.3\times10^4$~AU), the median separation in the degraded-resolution
Ophiuchus map ($2.6\times10^4$~AU) is consistent with the other two
clouds.   Imposing a uniform flux limit of
$75 \mathrm{mJy} \times(\frac{250\mathrm{pc}}{d})^2$ across all three clouds,
equivalent to the $5\sigma$ limit in the shallowest map (Perseus),
does not significantly change these results.  

We calculate the number of cores within one square parsec at each
point in the map, and take the peak value to be the peak number of
cores per square parsec.  The peak values are $12 ~\mathrm{pc}^{-2}$
in Serpens, $22 ~\mathrm{pc}^{-2}$ in Perseus, and $24
~\mathrm{pc}^{-2}$ in the original Ophiuchus map.  Differences in
linear resolution and completeness both have an effect in this case;
the peak number in Ophiuchus falls to 20~pc$^{-2}$ when using a
uniform flux limit, and to 12~pc$^{-2}$ in the degraded-resolution
map.  The peak number of  cores per parsec provides further evidence
that clustering is stronger  in Perseus than in the other two clouds.

%\subsection{Relationship to Cloud Column Density and Efficiency of Forming Cores}\label{threshsect}
\subsection{Relationship to Cloud Column Density}\label{threshsect}

In contrast to the extinction map, which is a measure of the  general
cloud (line-of-sight averaged) column density, the 1.1~mm  map is
sensitive only to regions of high volume density (see
\S~\ref{coresect}).  A comparison of the two tells us, therefore,
about the relationship between dense star-forming structures and the
column density of the larger-scale cloud.  A visual comparison of the
1.1~mm maps of each cloud with visual extinction maps derived from
the reddening of background stars (e.g. Figure~\ref{avfig}) suggests
that 1.1~mm cores are generally found in regions of the cloud with
high \av.

\begin{figure}
\plotone{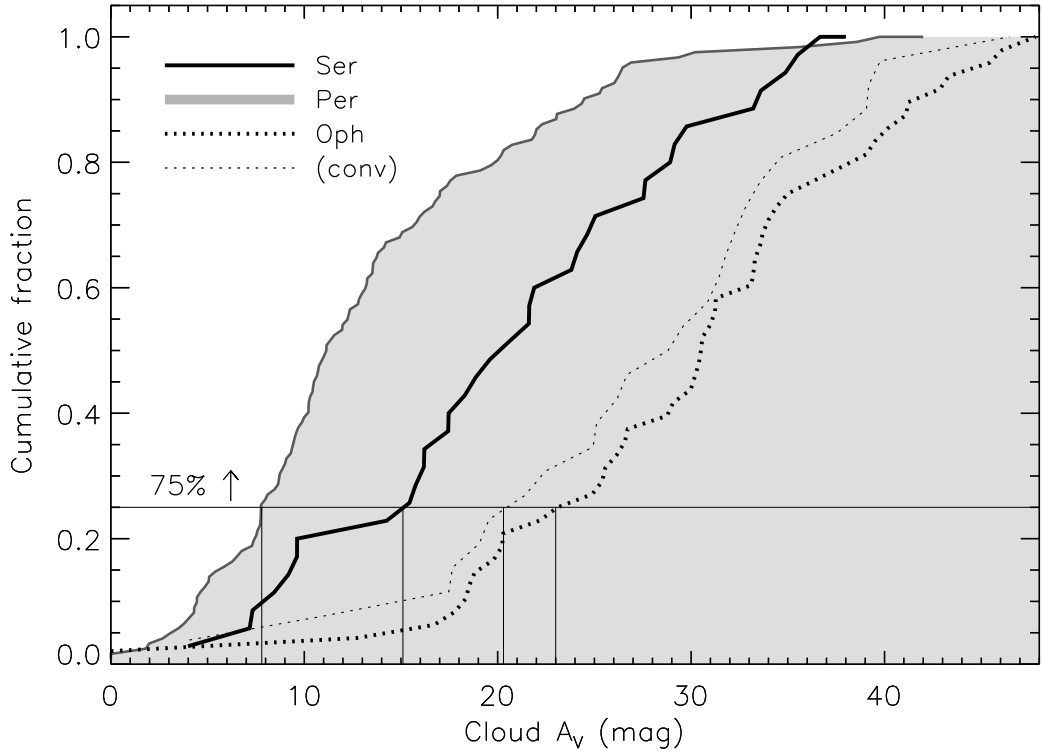}
\caption{ Cumulative fraction of 1.1~mm cores as a function of cloud
  \av\, for sources in Perseus, Serpens, and Ophiuchus.   The cloud
  \av\ is derived from the reddening of background stars using 2MASS
  and IRAC data as described in \S~\ref{ressect}.  Thin solid lines
  denote the \av\ levels above which 75\% of dense cores are found in each cloud.
  More than 75\% of 1.1~mm cores lie above $\av\sim8^m$ in Perseus,
  $\av\sim15^m$ in Serpens, and $\av\sim23^m$ in Ophiuchus.  Results
  for the degraded resolution Ophiuchus map are shown as well (thin
  dotted line), but do not vary substantially from the original
  Ophiuchus sample, with 75\% of cores above $\av\sim20^m$. 
\label{avcomp} }
\end{figure}

Figure~\ref{avcomp} quantifies the relationship between dense cores
and the surrounding cloud column density by plotting the cumulative
fraction of 1.1~mm cores in each cloud as a function of cloud \av.  In
all three clouds the majority of cores are found at high cloud column
density ($\av>7^m$).  The \av\ levels above which 75\% of cores are
found in each cloud are indicated by thin lines in  Figure~\ref{avcomp}:  75\% of
1.1~mm cores in Perseus, Serpens, and Ophiuchus are found at visual
extinctions of $\av \gtrsim 8^m$, $\av \gtrsim 15^m$, and $\av \gtrsim
23^m$, respectively.  Although there is not, in general,  a strict extinction
threshold for finding cores, below these \av\ levels the likelihood of finding a
1.1~mm core is very low.  
Only in Ophiuchus does there appear to be a true
\av\ threshold; only two cores are found at $\av < 17^m$ in this cloud.

The cumulative distribution for the degraded resolution Ophiuchus
sample is nearly identical to the original Ophiuchus sample, but
shifted to lower \av\ by 1--2~mag. For the degraded resolution map,
75\% of cores are found at $\av \gtrsim 20^m$. Based on this result and the
value for the original Ophiuchus sample, we adopt $\av\sim20-23^m$ for the 75\%
level in Ophiuchus.  Requiring that all cores in the Perseus and
Serpens sample meet the detection threshold for Ophiuchus
(approximately 110~mJy) changes the distributions negligably.  
Only for Serpens is the 75\% level increased slightly, from $\av \sim15^m$ to $16^m$.

Cloud to cloud differences could indicate variations in the core
formation process  with environment, differing degrees of
sub-structure in the clouds, or  varying amounts of foreground
extinction.   Note that in Papers I and II we found extinction
``thresholds'' for Perseus and Ophiuchus, using a different analysis
that looked at the probability of finding a 1.1~mm core as a function
of \av, of $\av \sim 5^m$ and $\av \sim 9^m$, respectively.  Those
values were derived using 2MASS-only \av\ maps, rather than  the c2d
extinction maps used here.

\citet{john04} have suggested an extinction threshold for forming
dense cores in Ophiuchus at $\av \sim 15^m$, based on the lowest \av\
at which SCUBA cores were observed to have sizes and  fluxes
consistent with those of stable Bonnor-Ebert spheres.  We find that
most Bolocam cores in Ophiuchus are found at even higher extinctions
(75\% at $\av>20-23^m$); the discrepancy likely arises from
differences in the extinction maps used.  \citet{john04} used a
2MASS-derived extinction  map, while the c2d extinction map used here
includes IRAC data as well and probes somewhat higher \av\ values.
\citet{hatch05} find no evidence for an extinction threshold in
Perseus using $850~\micron$ SCUBA data, but \citet{kirk06}, also using
SCUBA data, find a threshold of $\av \sim5^m$.  Despite differences in
detail, the relative 75\% \av\ levels found here (highest in Ophiuchus
and lowest in Perseus) are consistent with the relative values of
previous  extinction threshold measurements in Perseus and Ophiuchus.

An extinction threshold has been predicted by \citet{mckee89} for
photoionization-regulated star formation in magnetically supported
clouds.  In this model star formation is governed by ambipolar
diffusion, which depends on ionization levels in the cloud.  Core
collapse and star formation will occur only in shielded regions of a
molecular cloud where $\av \gtrsim 4-8^m$.   \citet{john04} note
that it is not clear how turbulent models of star formation would produce
an extinction threshold for star-forming objects, and
interpret such a threshold as evidence for magnetic support.

\subsection{Efficiency of Forming Cores}\label{efficsect}

Another interesting measure of global cloud conditions is the fraction
of  cloud mass contained in dense cores.  Table~\ref{avtab} lists the
cloud area and cloud mass within increasing \av\ contours, calculated
from the c2d visual extinction maps, as well as the total mass in
cores within the same extinction contour.   Cloud masses are
calculated from the c2d \av\ maps (\S~\ref{masssec},
Eq.~\ref{avmasseq}).  The total cloud mass within $\av>2^m$ is
3470~\msun for Serpens, 7340~\msun for Perseus, and 3570~\msun  for
Ophiuchus.

Within a given \av\ contour, the mass ratio is defined as the ratio of
total core mass to total cloud mass, and is a measure of the
efficiency of core formation at that \av\ level.  For example, the
mass ratio at the $\av=2^m$ contour is equivalent to the fraction of
total cloud mass that is contained in dense cores.  In each cloud
1.1~mm cores account for less than 5\% of the total cloud mass.  The
mass ratios at $\av=2^m$ are similar in all three clouds: 3.8\% in
Perseus, 2.7\% in Serpens, and 1.2\% in Ophiuchus.   If we restrict
ourselves to $\av>6^m$ in each cloud, which is reasonable given that
the Serpens map was only designed to cover  $\av>6^m$, the mass ratio
is still low in all three clouds ($<10$\%), and remains higher in
Perseus (7\%) than in Serpens (4\%) or Ophiuchus (2\%).  Low mass
ratios are consistent with measurements of the overall star-formation
efficiency of $1-2\%$, which suggests  that molecular cloud material
is relatively sterile \citep[e.g.][]{evans99,leis89}.   

\citet{john04} find a mass ratio of 2.5\% in a survey of approximately
$4$~deg$^2$ of Ophiuchus, quite similar to our result.  Note, however,
that the \citet{john04} core masses (and mass ratio) should be
multiplied by a factor of 1.5 to compare to our values, due to
differences in assumed values of $T_D$, $\kappa_{\nu}$, and $d$.  In a
similar analysis of 3.5 deg$^2$ in Perseus, also using SCUBA
$850~\micron$ data, \citet{kirk06} found a mass ratio for $\av>0$ of
only 1\%.  The difference between this result and our value arises
primarily from the smaller core masses of those authors, which  should
be multiplied by  2.5 to compare to ours.  

In all three clouds the mass ratio rises with increasing \av\ contour,
indicating that in high extinction regions a greater percentage of
cloud mass has been assembled into cores, consistent with the idea
that star formation is more efficient in dense regions.   Although
this is an intuitively obvious result, it is not a necessary one.  If,
for example, a constant percentage of cloud mass were contained in
dense cores at all column densities, there would be a large number of
dense cores lying in a low \av\  background.  On the other hand, a
molecular cloud might consist of large regions of uniformly high
extinction in which we would find no 1.1~mm cores because there is no
sub-structure, and millimeter cores require high density contrast (see
\S~\ref{coresect}).

At higher \av\ levels, mass ratios vary considerably from cloud to
cloud.   The mass ratio remains fairly low in Ophiuchus, with a
maximum value of 9\%  at $\av=30^m$.  In contrast, the mass ratio
rises rapidly in Serpens to 65\%  at $\av=30^m$, which may suggest
that Serpens has formed cores more efficiently  than Ophiuchus at high
\av.

\section{Summary}\label{sumsect}

This work completes a three-cloud study of the millimeter continuum
emission in Perseus, Ophiuchus, and Serpens.   We examine similarities
and differences in the current star formation activity within the
clouds using large-scale 1.1~mm continuum maps completed with Bolocam
at the CSO.   In total, our surveys cover nearly 20 deg$^2$ with a
resolution of 31\arcsec\  (7.5 deg$^2$ in  Perseus, 10.8 deg$^2$ in
Ophiuchus, and 1.5 deg$^2$ in Serpens),  and  we have assembled a
sample of 200 cores (122 in Perseus,  44 in Ophiuchus, and 35 in
Serpens).   Point mass detection limits vary from approximately 0.1 to
0.2 \msun\ depending  on the cloud.  The results presented here
provide an unprecedented global picture of star formation in three
clouds spanning a range of diverse environments.

These Bolocam 1.1~mm observations naturally select dense cores with
$n\gtrsim 2\times 10^4$~cm$^{-3}$ and density contrast compared to  the
background cloud of at least $30-100$.  We test instrumental biases
and the effects of cloud  distance by degrading the resolution of the
Ophiuchus map  to match the distance of Perseus and Serpens.  We find
that linear resolution strongly biases measured  linear deconvolved
source sizes and mean densities,  but not the mass distribution slope.
Angular deconvolved sizes are less strongly affected by cloud
distance.

Rather than a true physical difference, the small mean linear
deconvolved sizes in  Ophiuchus ($0.8 \times 10^4$~AU) compared to
Perseus ($1.5 \times 10^4$~AU) and Serpens ($1.2 \times 10^4$~AU) are
likely a result of observing sources with power law density profiles,
which do not have a well defined size, at a distance of 125~pc in
Ophiuchus versus 250~pc in Perseus and 260~pc in Serpens.  The
observed mean angular deconvolved sizes and axis ratios in each cloud
suggest average power law indices ranging from $p=1.4$ to 1.7 (Y03).

Sources in Perseus exhibit the largest range in sizes, axis ratios,
and densities, whereas sources in both Serpens and Ophiuchus display a
fairly narrow range of sizes for a large range of masses.   We suggest
that this is indicative of a greater variety of physical conditions in
Perseus, supported by the fact that Perseus contains both dense
clusters of millimeter sources and more isolated distributed objects.
A wide range in angular deconvolved  sizes may also imply a range in
the power law index of source profiles in Perseus (Y03). 

The slope of the clump mass distribution for both Perseus and
Ophiuchus is $\alpha=2.1$, marginally different than the Serpens slope
of $\alpha=1.6$.  Only Perseus and Ophiuchus are consistent within the
substantial errors with the stellar initial mass function ($\alpha
\sim 2.5-2.8$) and with the slope predicted for turbulent
fragmentation ($\alpha=2.3$) by \citet{padoan02}.

Turbulent fragmentation simulations by BP06 predict that higher cloud
Mach numbers should result in a large number of low mass  cores, and
low Mach numbers in a smaller number of higher mass cores.  Given the
measured Mach numbers of $\mathcal{M}=4.9$ in Serpens, 3.6 in Perseus
and 2.3 in Ophiuchus, our observed core mass distribution (CMD) shapes
are inconsistent with the  turbulent fragmentation prediction from
BP06.  We cannot rule out a turbulent fragmentation scenario, however,
due to uncertainties in the observations and in  our assumptions.

We argue that clustering of 1.1~mm sources remains stronger out to
larger scales in Perseus, based on the slope of the two-point
correlation  function (-1.5 in Serpens and Ophiuchus, and -1.2 in
Perseus).  This result is supported by the fact that the peak number
of cores per square parsec is larger in Perseus
($22~\mathrm{pc}^{-2}$) than in Serpens ($12 ~\mathrm{pc}^{-2}$) or
the degraded-resolution Ophiuchus map ($12 ~\mathrm{pc}^{-2}$).

Finally, we examine relationship between dense cores and the local
cloud column density, as measured by visual extinction (\av).
Extinction thresholds for star formation have been suggested based on
both theory and observation \citep{mckee89, john04}.  Although in
general we do not observe a strict \av\ threshold, dense 1.1~mm cores
do  tend to be found at high \av:  75\% of cores in Perseus are found
at  $\av\gtrsim 8^m$, in Serpens at $\av\gtrsim 15^m$, and in
Ophiuchus at  $\av\gtrsim 20-23^m$.  Our results confirm that forming
dense cores in molecular clouds is a very inefficient process, with
1.1~mm cores accounting for less than 10\% of the total cloud mass in
each cloud.  This result is consistent with measurements of low star
formation efficiencies of a few percent from studies of the stellar
content of molecular clouds \citep[e.g.][]{evans99}.

While millimeter-wavelength observations can provide a wealth of
information about the detailed properties of star forming cores as
well as insight into the large scale physical properties of molecular
clouds, they do not tell a complete story.  Detecting and
understanding the youngest embedded protostars currently forming
within those cores requires information at mid- to far-infrared
wavelengths.   The Bolocam maps for all three clouds presented here
are coordinated to cover the same regions as the c2d \textit{Spitzer}
Legacy IRAC and MIPS maps of Serpens, Perseus, and Ophiuchus.
Combining  millimeter and \textit{Spitzer} data for these clouds will
allow us to separate starless cores from cores with embedded
luminosity sources and to better understand the evolution of cores
through the early Class 0  and Class I protostellar phases.  Synthesis
and analysis of the combined data set is currently  underway (M. Enoch
et al. 2007, in preparation).

\acknowledgments

The authors are grateful to Paul Harvey, Yancy Shirley, and the 
anonymous referee for comments that significantly improved this work.
We would like to thank members of the Bolocam team for instrumental
and software support, including James Aguirre, Jack Sayers, Glenn Laurent, 
and Sunil Golwala.  We also thank the Lorentz Center in Leiden for
hosting several meetings that contributed to this paper.  Support for
this work, part of the Spitzer Legacy Science Program, was provided by
NASA through contracts 1224608 and 1230782 issued by the Jet
Propulsion Laboratory, California Institute of Technology, under NASA
contract 1407.  Bolocam was built and commissioned under grants
NSF/AST-9618798 and NSF/AST-0098737.  KEY was supported by
NASA under Grant NGT5-50401, issued
through the Office of Space Science.  Additional support came from
NASA Origins grant NNG04GG24G to NJE and NSF grant AST 02-06158 to JG.
MLE acknowledges support of a Moore Fellowship and an NSF Graduate 
Research Fellowship.

%%%%%%%%%%%%%% References %%%%%%%%%%%%%%%%%%%%%%

%%%%%%%%%%%%%%%%%% Tables %%%%%%%%%%%%%%%%%%%%%

\clearpage
\input{tab1}

        \clearpage
%        \LongTables % optionally
        \begin{landscape}
     \input{tab2}
         \clearpage
        \end{landscape}
%\clearpage
%\thispagestyle{empty}
%\input{tab2}

\clearpage
\input{tab3}			

\clearpage

%%%%%%%%%%%%%%%%%%% Figures %%%%%%%%%%%%%%%%%%

\epsscale{0.65}

\epsscale{0.9}

\epsscale{0.9}

\epsscale{0.95}
\notetoeditor{Figure 4 should be full-page, and color in the printed version.}  

\epsscale{0.65}

\epsscale{1.}

\epsscale{0.75}

\epsscale{1.}

\epsscale{1.}

\epsscale{0.9}

\epsscale{0.8}

\epsscale{1.0}

\end{document}